\newif\ifarxiv
\arxivtrue

\PassOptionsToPackage{svgnames,dvipsnames}{xcolor}
\ifarxiv
  \documentclass[sigconf,nonacm,screen,natbib=false]{acmart}
\else
  \documentclass[sigconf,nonacm,natbib=false]{acmart}
\fi

\ifarxiv
\else
\usepackage{pvldb}%

\renewcommand\vldbdoi{XX.XX/XXX.XX}
\renewcommand\vldbpages{XXX-XXX}
\renewcommand\vldbavailabilityurl{https://github.com/ngmarchant/actis} 
\fi

\RequirePackage[
  datamodel=acmdatamodel,
  style=acmnumeric,
  natbib=true,
  sortcites=true
  ]{biblatex}

\usepackage{tikz}
\usetikzlibrary{shapes.geometric, arrows.meta, positioning, calc, fit, backgrounds}
\usepackage{algorithm}
\usepackage{algpseudocode}
\usepackage{mathtools}
\usepackage{import} %
\usepackage{pgf} %
\usepackage{placeins} %
\usepackage{enumitem}
\usepackage[capitalize]{cleveref}

\crefname{section}{Sec.}{Secs.}
\Crefname{section}{Section}{Sections}
\crefname{appendix}{App.}{Apps.}
\Crefname{appendix}{Appendix}{Appendices}
\crefname{algorithm}{Alg.}{Algs.}
\Crefname{algorithm}{Algorithm}{Algorithms}
\crefname{proposition}{Proposition}{Propositions}
\Crefname{proposition}{Proposition}{Propositions}

\newcommand{\myparagraph}[1]{\addvspace{3pt plus 1pt minus 1pt}\emph{\sffamily #1}}
\newcommand{\wmax}{w_{\max}}
\newcommand{\indicator}[1]{\mathbf{1}_{\{#1\}}}

\newcommand{\D}{\mathcal{D}}

\newcommand{\scrX}{\mathcal{X}}

\newcommand{\grid}{G}
\newcommand{\lo}[1]{\check{#1}}
\newcommand{\up}[1]{\hat{#1}}
\DeclareMathOperator{\oracle}{O}
\DeclareMathOperator{\proxy}{A}
\DeclareMathOperator{\cascade}{C}
\DeclareMathOperator*{\E}{E}
\DeclareMathOperator{\var}{Var}

\DeclareMathOperator{\Recall}{Recall}
\DeclareMathOperator{\Precision}{Precision}
\DeclareMathOperator{\TPR}{TPR}
\DeclareMathOperator{\FPR}{FPR}
\DeclareMathOperator{\FNR}{FNR}
\DeclareMathOperator{\Prevalence}{\pi}

\DeclarePairedDelimiter{\abs}{\lvert}{\rvert}

\newcommand{\badvalue}[1]{\textcolor{Maroon}{#1}}

\newenvironment{romanenum}{%
  \begin{list}{}{%
    \usecounter{enumi}%
    \renewcommand{\makelabel}[1]{\llap{\textup{(\roman{enumi})}\hskip 0.5em}}%
    \setlength{\labelwidth}{0pt}%
    \setlength{\leftmargin}{0.8\leftmargini}%
  }%
}{%
  \end{list}%
}

\begin{document}

\title{ACTIS: Adaptive Cascade Tuning for Semantic Filters with Statistical Guarantees}

\author{Neil Marchant}
\affiliation{%
  \institution{University of Melbourne}
  \country{}
}
\email{nmarchant@unimelb.edu.au}

\author{Renata Borovica-Gajic}
\affiliation{%
  \institution{University of Melbourne}
  \country{}
}
\email{rborovica@unimelb.edu.au}

\author{Benjamin Rubinstein}
\affiliation{%
  \institution{University of Melbourne}
  \country{}
}
\email{brubinstein@unimelb.edu.au}

\author{Olga Ohrimenko}
\affiliation{%
  \institution{University of Melbourne}
  \country{}
}
\email{oohrimenko@unimelb.edu.au}
\settopmatter{authorsperrow=4}
\begin{abstract}
Emerging semantic database systems use Large Language Models (LLMs) to evaluate natural language predicates during query processing. However, invoking an expensive frontier model for every tuple is prohibitively costly and induces high latency. 
Model cascades mitigate these limitations by using a cheap proxy to score all tuples, escalating only uncertain predictions to an expensive LLM oracle. 
To be deployed reliably, cascades are tuned on a sample of oracle-labeled tuples to satisfy precision and recall targets with high probability. 
Our comprehensive audit reveals that several existing tuning methods frequently violate these guarantees: state-of-the-art techniques are certified unsound on up to 90.5\% of benchmarks and can miss accuracy targets by more than 70~percentage points.

To improve efficiency and ensure soundness guarantees for semantic databases, we design ACTIS, an adaptive cascade tuning framework. 
ACTIS acquires oracle labels incrementally using anytime-valid tests with finite-sample safeguards, avoiding the over- or under-sampling pitfalls of a fixed pilot sample. 
A decoupled fixed-sequence search explores fine-grained threshold grids without conservative multiple-testing corrections, while a variance-minimizing importance sampling distribution reduces labels spent on uninformative negatives for selective queries. 
A value-of-information stopping rule halts tuning when additional oracle labeling costs exceed expected savings across the remaining workload. 
Across 42 benchmarks with cloud and local proxies, ACTIS matches or outperforms the only sound baseline on 39 of 42 settings, reducing oracle calls by up to 5.1$\times$.
\end{abstract}

\maketitle

\ifarxiv
\noindent\textbf{Artifact Availability.} The source code and data have been made available at \url{https://github.com/ngmarchant/actis}.
\medskip
\else
\vldbtopmatter
\fi

\section{Introduction} \label{sec:intro}

Predicate evaluation is a foundational primitive of database management systems. 
In traditional query engines, declarative queries apply attribute-based predicates over structured columns (e.g., $\texttt{AGE} \ge 18 \land \texttt{STATE} = \text{`CA'}$) to filter tuples early in query execution plans. 
With the rise of foundation models and the rapid influx of unstructured data---such as text documents, customer communications, and multi-modal records---a new class of \emph{semantic data processing systems} has emerged. 
Systems such as LOTUS~\citep{patel2025semantic}, Palimpzest~\citep{liu2025palimpzest}, and DocETL~\citep{shankar2025docetl}, alongside enterprise engines like Google BigQuery~\citep{he2025sql} and Snowflake Cortex AISQL~\citep{liskowski2026cortex}, extend relational engines with declarative \emph{semantic operators} \citep{dai2024uqe,dorbani2025beyond,lin2025querying,sun2025quest,zhu2026beyond,kumarasinghe2026ipdb,qi2026sema}.
In particular, the \emph{semantic filter} evaluates natural-language conditions over tuple attributes expressed via prompt templates---for example:
\begin{equation*}
    \Call{SemFilter}{\text{``Is \{\texttt{summary}\} an accurate synopsis of \{\texttt{full\_text}\}?''}},
\end{equation*}
where $\texttt{summary}$ and $\texttt{full\_text}$ are tuple attributes.

\begin{figure}[t]
    \centering
    \resizebox{0.85\linewidth}{!}{\input{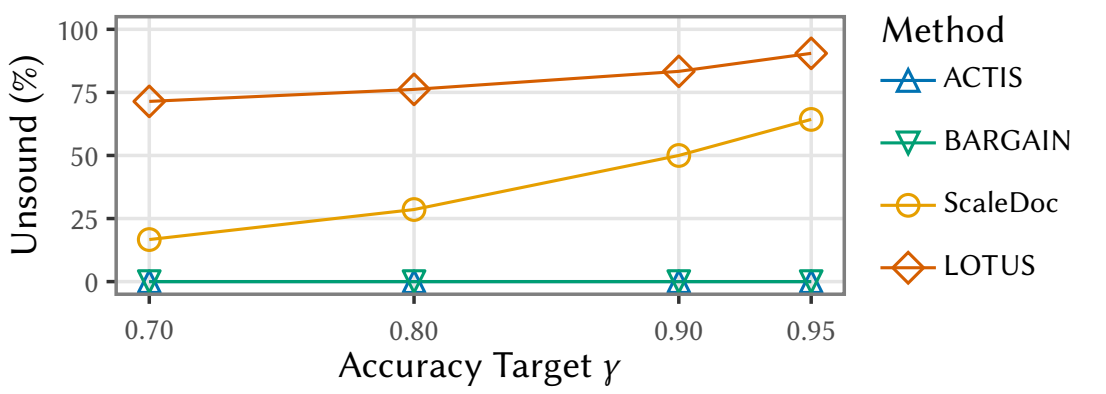}}
    \addvspace{-6pt}
    \caption{Percentage of 42 benchmarks failing statistical accuracy guarantees at $\delta = 0.05$ versus accuracy target $\gamma$. 
    LOTUS~\cite{patel2025semantic} and ScaleDoc~\cite{zhang2026scaledoc} fail on up to 90.5\% and 64.3\% of benchmarks, respectively, while ACTIS and BARGAIN~\cite{shankar2026bargainpr} achieve zero unsound runs.%
    }
    \label{fig:teaser-soundness}
\end{figure}

The most direct way to execute a semantic filter is to submit every tuple to an expensive frontier Large Language Model (LLM), which acts as an \emph{oracle}.
Although AI model providers offer batch APIs that discount high-throughput requests~\citep{openai2026pricing}, evaluating an oracle LLM across millions of records remains prohibitively expensive and introduces high query latency.
Filtering tuples early drastically reduces intermediate data volumes for downstream join, aggregation, and synthesis operators; consequently, optimizing semantic filters is critical to overall semantic query execution~\citep{patel2025semantic}.

To reduce execution costs, recent work leverages \emph{model cascades}~\citep{viola2001rapid,kang2020approximate,patel2025semantic,zeighami2025cut,zhang2026scaledoc,sanmartino2026stretto}. 
In a two-threshold cascade, a fast, lightweight proxy model---such as a small language model or an on-the-fly classifier trained over vector embeddings~\citep{chung2026cost}---scores each tuple with a confidence value in $[0, 1]$.
Tuples with proxy confidence exceeding an upper threshold $\up{\tau}$ are predicted directly as \texttt{True} without calling the oracle; tuples falling below a lower threshold $\lo{\tau}$ are predicted directly as \texttt{False}; only ambiguous intermediate tuples are escalated to the oracle.
To be deployed reliably within a semantic data processing system, such cascades must guarantee output fidelity. Operationally, this means satisfying user-specified precision and recall targets $\gamma_P, \gamma_R \in (0, 1)$ with respect to the oracle with probability at least $1 - \delta$, where the probability is over the randomness in the cascade's tuning procedure.

Several pioneering cascade frameworks---notably LOTUS~\citep{patel2025semantic}, ScaleDoc~\citep{zhang2026scaledoc}, and SUPG-IT~\citep{liskowski2026streaming}---claim to provide such statistical guarantees.
However, our comprehensive empirical audit reveals a concerning reality: \emph{the accuracy guarantees claimed by existing semantic filter cascades are frequently unsound in practice}.
As illustrated in \cref{fig:teaser-soundness}, under a standard error budget $\delta = 0.05$ with accuracy target $\gamma = 95\%$, LOTUS's guarantees are certified unsound---meaning its empirical failure rate statistically exceeds the error budget~$\delta$---on 90.5\% of benchmarks, while ScaleDoc's guarantees are certified unsound on 64.3\% of benchmarks.
Moreover, the extent of these violations is severe: LOTUS falls short of the 95\% target by more than 76 percentage points in the worst case, and ScaleDoc falls short by 86 percentage points.

In database query processing, an unsound accuracy guarantee is arguably worse than having no guarantee at all. 
Guarantees act as an SLA: users rely on them to ensure that approximate filtering will neither drop critical matching records (violating recall) nor pollute downstream queries with irrelevant data (violating precision).
When an algorithm violates this contract, its cost savings are misleading, since any filter can reduce cost by silently dropping relevant tuples.

Why do existing tuning algorithms fail to uphold this contract?
Our analysis reveals that these failures stem from compounding theoretical oversights and implementation gaps \ifarxiv~(\cref{app:soundness-analysis})\else~(see App.~C of our extended paper~\citep{actis_full})\fi.
Subtle flaws---ranging from uncorrected multiple hypothesis testing across grids of thresholds to fragile approximations in skewed data regimes---bias cascades toward overly aggressive thresholds whose true failure rates substantially exceed the nominal error budget~$\delta$.

While single-threshold cascade techniques like BARGAIN~\citep{zeighami2025cut} establish finite-sample validity, the published literature lacks sound algorithms for two-threshold cascades. 
A recent, unpublished extension by BARGAIN's authors, \emph{BARGAIN-PR}~\citep{shankar2026bargainpr},\footnote{%
    For brevity we refer to the two-threshold BARGAIN-PR extension simply as ``BARGAIN'' throughout our evaluation.
}
extends their framework to handle two-threshold cascades under uniform sampling, but suffers from sample inefficiency on imbalanced workloads---such as scanning thousands of records just to observe rare positives. 
Moreover, it lacks cost-aware stopping criteria that balance expenditure for tuning against expenditure for the remaining workload.

To resolve this conflict between efficiency and correctness, we formulate cascade tuning as an \emph{online, cost-aware query optimization} problem.
We present \textbf{ACTIS} (\textbf{A}daptive \textbf{C}ascade \textbf{T}uning via \textbf{I}mportance \textbf{S}ampling), a framework for tuning two-threshold model cascades that ensures statistical soundness while minimizing the total number of oracle model invocations. 
ACTIS introduces four key technical contributions:

    \myparagraph{(1)~Decoupled Fixed-Sequence Testing.} 
    A naive search over a 2D~grid of lower and upper thresholds $\lo{\grid} \times \up{\grid}$ requires severe Bonferroni corrections, requiring impractically large oracle samples for tuning. 
    By proving that precision and recall margins exhibit monotonicity, we decouple the 2D search into two sequential 1D stages. 
    Fixed-sequence testing~\citep{maurer1995multiple} evaluates candidate thresholds in monotonic order with no multiplicative penalty over the upper threshold grid, enabling dense search without statistical power loss.

    \myparagraph{(2)~Anytime-Valid Testing with Safeguards.} 
    Rather than relying on a predetermined pilot sample, ACTIS evaluates candidate thresholds sequentially using anytime-valid test processes~\citep{howard2021time,ramdas2023game,waudby2024time}. 
    To prevent failure modes of asymptotic Gaussian mixture tests on small or skewed samples, ACTIS integrates three finite-sample safeguards: a safe prior configuration, an empirical variance accumulation check, and a binomial check that prevents premature stopping on early success streaks.

    \myparagraph{(3)~Variance-Minimizing Importance Sampling.} 
    Uniform sampling wastes oracle budget by scanning uninformative true negatives when target positives are rare. 
    We derive the closed-form instrumental distribution that directly minimizes the total asymptotic variance of the empirical estimators for true positive rate, false positive rate, and positive prevalence, concentrating oracle calls on informative samples for tuning.

    \myparagraph{(4)~Adaptive Stopping.} 
    Balancing tuning cost against expected savings across the remaining workload requires knowing when additional oracle labels cease to be cost-effective. 
    Drawing on the Value of Information framework~\citep{howard1966information}, ACTIS constructs optimistic threshold bounds and halts tuning when the expected reduction in oracle invocations across the remaining workload falls below the marginal acquisition cost of the next batch.

\smallskip
We evaluate ACTIS across 42 benchmark scenarios spanning 23 datasets and queries derived from prior literature~\citep{kang2020approximate,zeighami2025cut,zhang2026scaledoc}, evaluating both frontier cloud models (GPT-5.6) and local open-weight models (Llama-3.2-1B). 
Across all 42 benchmarks, ACTIS strictly adheres to the user-specified error budget $\delta$, with zero certified unsound runs. 
Compared to BARGAIN---the only baseline that maintains statistical validity---ACTIS matches or reduces oracle call rates on 39 of 42 benchmarks, reducing oracle invocations by up to $5.1\times$ on challenging workloads. 
Furthermore, our ablation studies show that variance-optimal importance sampling delivers its greatest benefits when proxy models are well-calibrated and query predicates are highly selective.
Through these findings, we call on the database community to prioritize auditing of statistical guarantees, ensuring that cost optimizations in semantic query processing remain dependable in production.

\section{Preliminaries} \label{sec:prelim}

In this section, we formalize the problem of tuning model cascades for semantic filters and introduce the foundational principles of anytime-valid hypothesis testing.

\subsection{Problem Formulation} \label{sec:problem}

Let $\scrX$ denote a domain of data records (e.g., text passages or document pairs). 
A workload consists of a finite table $\D \in \scrX^N$ containing $N$ records. 
We denote by $p$ the nominal uniform distribution over the workload table, such that $p(x) = 1/N$ for all $x \in \D$.

\myparagraph{Oracle and Proxy Models.}
A semantic filter applies a predicate to each record in $\D$. 
Ideally, this predicate is evaluated by an accurate, highly capable model, which we refer to as the \emph{oracle model} $\oracle$. 
For each record $x \in \D$, evaluating the oracle yields a binary label $Y \sim \oracle(x) \in \{0, 1\}$. 
This formulation accommodates both deterministic oracles (such as greedy LLM decoding) and stochastic oracles (such as reasoning models, where $\E[Y | x] \in [0, 1]$).
Evaluating the oracle incurs high latency and monetary cost. 
To alleviate this burden, a fast, lightweight \emph{proxy model} $\proxy \colon \scrX \to [0, 1]$ assigns a confidence score $\proxy(x)$ to each record $x \in \D$.

\myparagraph{Two-Threshold Model Cascade.}
To balance efficiency and fidelity, we approximate the oracle using a two-threshold model cascade $\cascade \colon \scrX \to \{0, 1\}$, parameterized by a lower threshold $\lo{\tau} \in [0, 1]$ and an upper threshold $\up{\tau} \in [\lo{\tau}, 1] \cup \{\infty\}$:
\begin{equation*}
    \cascade(x; \lo{\tau}, \up{\tau}) = \begin{cases}
        1,              & \proxy(x) \ge \up{\tau}, \\
        0,              & \proxy(x) < \lo{\tau}, \\
        \oracle(x),     & \lo{\tau} \le \proxy(x) < \up{\tau}.
    \end{cases}
\end{equation*}
The cascade predicts \texttt{True} (1) for records whose proxy scores are~$\ge \up{\tau}$, predicts \texttt{False} (0) for records whose proxy scores are $< \lo{\tau}$, and escalates intermediate records to the expensive oracle.
The primary cost metric of the cascade is the \emph{oracle call rate} $r(\lo{\tau}, \up{\tau})$, defined as the fraction of the workload deferred to the oracle:
\begin{equation}
    r(\lo{\tau}, \up{\tau}) = \frac{1}{N} \sum_{x \in \D} \indicator{\lo{\tau} \le \proxy(x) < \up{\tau}}.
    \label{eqn:ocr}
\end{equation}

\myparagraph{Workload Quality Metrics.}
Evaluating the quality of the cascade relative to the oracle requires tracking the positive prevalence, true positive rate, and false positive rate across the workload $\D$:
\begin{align}
    \Prevalence &= \E_{X \sim p, Y \sim \oracle(X)}[Y], \label{eqn:pp} \\
    \TPR(\lo{\tau}) &= \E_{X \sim p, Y \sim \oracle(X)}\bigl[Y \indicator{\proxy(X) \ge \lo{\tau}}\bigr], \label{eqn:tpr} \\
    \FPR(\up{\tau}) &= \E_{X \sim p, Y \sim \oracle(X)}\bigl[(1 - Y) \indicator{\proxy(X) \ge \up{\tau}}\bigr]. \label{eqn:fpr}
\end{align}
The cascade recall and precision over $\D$ are then given by:
\begin{align}
    \Recall(\lo{\tau}) &= \frac{\TPR(\lo{\tau})}{\Prevalence}, \label{eqn:recall} \\
    \Precision(\lo{\tau}, \up{\tau}) &= \frac{\TPR(\lo{\tau})}{\TPR(\lo{\tau}) + \FPR(\up{\tau})}. \label{eqn:precision}
\end{align}
When the oracle is deterministic, expectations under $X \sim p, Y \sim \oracle(X)$ collapse to empirical averages over $\D$.

\myparagraph{Cascade Tuning with Accuracy Guarantees.}
Given user-specified precision and recall targets $\gamma_P, \gamma_R \in (0, 1)$ and an error budget $\delta \in (0, 1)$, our objective is to find thresholds $(\lo{\tau}^\star, \up{\tau}^\star)$ that minimize the oracle call rate while guaranteeing that both targets are satisfied simultaneously with probability at least $1 - \delta$:
\begin{equation}
\begin{aligned}
    &\min_{\lo{\tau} \in [0, 1], \up{\tau} \in [\lo{\tau}, 1] \cup \{\infty\}} \quad r(\lo{\tau}, \up{\tau}) \\
    &\text{s.t.} \quad \Pr\bigl[\Precision(\lo{\tau}, \up{\tau}) \ge \gamma_P \land \Recall(\lo{\tau}) \ge \gamma_R\bigr] \ge 1 - \delta,
\end{aligned}
\label{eqn:opt-problem}
\end{equation}
where the probability is taken strictly with respect to the randomness in the cascade tuning algorithm, as described below.

\myparagraph{Execution Model.}
Following prior cascade frameworks~\citep{kang2020approximate,patel2025semantic,zeighami2025cut,zhang2026scaledoc}, we assume proxy scores $\mathcal{S} = \{\proxy(x) : x \in \D\}$ are computed offline for all $N$ records prior to tuning, and that oracle labels acquired during tuning are cached and reused when the tuned cascade is deployed over the remaining workload. 
Unlike static methods~\citep{patel2025semantic,zhang2026scaledoc} that draw a single predetermined pilot sample, adaptive methods (\citealp{shankar2026bargainpr} and ACTIS) query the oracle sequentially in batches until statistical criteria are met.

\subsection{Anytime-Valid Testing} \label{sec:anytime-tests}

To certify that candidate thresholds satisfy the precision and recall constraints in~\cref{eqn:opt-problem}, the tuner must test statistical hypotheses over oracle-labeled samples. 
In an adaptive setting, batches of labeled data arrive sequentially. 
Evaluating standard fixed-sample confidence bounds (such as Hoeffding or empirical Bernstein bounds) repeatedly after each batch introduces the well-known \emph{peeking problem}: repeatedly monitoring a fixed-sample test inflates the Type~I error rate, causing the cumulative probability of a false rejection to approach 1 as sampling continues \citep{johari2017peeking}.

To evaluate hypotheses across sequential batches without error inflation, we leverage \emph{anytime-valid testing}~\citep{ramdas2023game}.
Let $Z_1, Z_2, \dots$ be a sequence of observations with conditional expectations $\mu = \E[Z_t \mid Z_1, \dots, Z_{t-1}]$. 
Suppose we wish to test the one-sided null hypothesis that $\mu$ is strictly below a target margin $m$, $H_0 \colon \mu < m$ versus $H_1 \colon \mu \ge m$.
An anytime-valid test tracks a non-negative process $(M_t)_{t=0}^\infty$ starting at $M_0 = 1$ and rejects $H_0$ at the stopping time $T = \inf\{t \ge 1 \colon M_t \ge 1/\alpha\}$, guaranteeing Type~I error control uniformly over all stopping times:
\begin{equation}
    \Pr_{H_0}\bigl(\exists t \ge 1 \colon M_t \ge 1/\alpha\bigr) \le \alpha.
    \label{eqn:anytime-guarantee}
\end{equation}
Below, we review two anytime-valid testing frameworks: non-asymptotic betting martingales and asymptotic mixture processes.

\subsubsection{Non-Asymptotic Testing via Betting Martingales} \label{sec:non-asymptotic-test}
For a true anytime-valid test that guarantees validity across all $t \ge 1$, we can construct $(M_t)_{t=0}^\infty$ as a \emph{test supermartingale} under $H_0$, satisfying $\E[M_t \mid Z_1, \dots, Z_{t-1}] \le M_{t-1}$ for all $t \ge 1$ under $H_0$.
By Ville's inequality~\citep{ville1939etude}, the maximum of a non-negative supermartingale satisfies $\Pr_{H_0}(\exists t \ge 1 \colon M_t \ge 1/\alpha) \le \E[M_0] / (1/\alpha) = \alpha$, ensuring \cref{eqn:anytime-guarantee} holds.

\begin{figure*}[t]
    \centering
    \resizebox{0.97\linewidth}{!}{\begin{tikzpicture}[
    node distance=0.4cm and 1.1cm,
    font=\sffamily\footnotesize,
    >={Stealth[length=4pt, width=3pt]},
    line width=0.65pt,
    panel/.style={
        rounded corners=4pt, draw=black!20, line width=0.6pt, inner sep=8pt, minimum height=3.4cm, yshift=0.2cm
    },
    stageheader/.style={
        font=\sffamily\bfseries\small, text=black, align=center
    },
    box/.style={
        rectangle, rounded corners=2.5pt,
        align=center, inner sep=4pt, font=\sffamily\footnotesize,
        minimum width=1.7cm, minimum height=0.8cm
    },
    databox/.style={
        box, draw=blue!70!black, fill=white, line width=0.7pt
    },
    procbox/.style={
        box, draw=blue!70!black, fill=white
    },
    modelbox/.style={
        box, draw=black!80, fill=black!8, line width=0.8pt, font=\sffamily\footnotesize
    },
    tunebox/.style={
        box, draw=violet!80!black, fill=white
    },
    decision/.style={
        diamond, aspect=1.5,
        draw=green!50!black, fill=green!12,
        font=\sffamily\bfseries\footnotesize, align=center, inner sep=2pt
    },
    deploybox/.style={
        box, draw=teal!80!black, fill=white
    },
    annot/.style={font=\sffamily\footnotesize, text=black!80, align=center}
]

\node (dataset) [databox] {Dataset\\$\D$};
\node (scores)  [procbox, below=of dataset] {Proxy Scores\\$\mathcal{S} = \{\proxy(x)\}$};
\node (proxy)   [modelbox, left=of scores] {Proxy Model\\$\proxy$};
\node (dist)    [procbox, right=of scores] {Sampling Dist.\\$q$};
\node (grids)   [procbox, above=of dist] {Threshold Grids\\$\lo{\grid}, \up{\grid}$};

\draw[->] (dataset.south) -- (scores.north);
\draw[->] (proxy.east)   -- (scores.west);
\draw[->] (scores.east) -- (dist.west);
\draw[->] ($(scores.east)+(0,0.2)$) -- ++(0.3,0.0) |- (grids.west);
\draw[->] (grids.south) -- (dist.north);

\node (sampler)    [tunebox, right=of dist] {Sampler};
\node (oracletune) [modelbox] at (grids -| sampler) {Oracle\\$\oracle$};
\node (cache)      [tunebox, right=of oracletune] {Labeled Cache\\$\mathcal{L}$};
\node (tune) [tunebox, right=of cache] {Tune Thresholds};
\node (stop) [decision] at (sampler -| tune) {Stop?};

\draw[->] (dist.east)       -- (sampler.west);
\draw[->] (sampler.north)    -- node[right, annot] {Unlabeled batch} (oracletune.south);
\draw[->] (oracletune.east) -- node[above, annot] {Labeled\\batch} (cache.west);
\draw[->] (cache.east)       -- (tune.west);
\draw[->] (grids.north)      -- ++(0,0.2) -|(tune.north);
\draw[->] (tune.south)       -- node[right, annot] {$(\lo{\tau}^\star, \up{\tau}^\star)$} (stop.north);
\draw[->] (stop.west) -- node[below, annot, align=right, pos=0.0, anchor=north east] {\textbf{No}} (sampler.east);

\node (remwork) [databox, right=of tune] {Remaining Dataset\\$\D \setminus \mathcal{L}$};
\node (cascade) [deploybox] at (stop -| remwork) {Model Cascade\\$\cascade(\cdot; \lo{\tau}^\star, \up{\tau}^\star)$};
\node (result)       [databox, right=of remwork] {Filtered Result\\Table};
\node (oracledeploy) [modelbox] at (cascade -| result) {Oracle\\$\oracle$};

\draw[->] (remwork.south) -- node[right, annot] {Records} (cascade.north);
\draw[->] (cascade.east) -- node[below, annot] {Uncertain} (oracledeploy.west);
\draw[->] (oracledeploy.north) -- (result.south);

\draw[->] ($(cascade.east)+(0,0.2)$) -- ++(0.3,0.0) 
    |- node[pos=0.25, right, annot] {Certain} ($(result.west)-(0,0.2)$);

\draw[->] (stop.east) -- node[below, annot, align=left, anchor=north west, pos=0.0] {\textbf{Yes}} (cascade.west);
\draw[->] (cache.north) -- ++(0,0.30) -| (result.north);
\draw[->] (scores.south) -- ++(0,-0.2) -| (cascade.south);

\begin{scope}[on background layer]
    \node[panel, fill=blue!3, fit=(dataset) (scores) (proxy) (dist) (grids)] (stage1) {};
    \node[panel, fill=violet!3, fit=(sampler) (oracletune) (cache) (tune) (stop)] (stage2) {};
    \node[panel, fill=teal!3, fit=(cascade) (remwork) (oracledeploy) (result)] (stage3) {};
\end{scope}

\node[stageheader] at ($(stage1.north)+(0,-0.3)$) {I. Initialization};
\node[stageheader] at ($(stage2.north)+(0,-0.3)$) {II. Adaptive Tuning Loop};
\node[stageheader] at ($(stage3.north)+(0,-0.3)$) {III. Workload Deployment};

\end{tikzpicture}}
    \caption{Overview of the three-stage ACTIS framework: (I)~proxy scoring, threshold grid construction, and importance sampling initialization; (II)~sequential batch acquisition, anytime-valid threshold tuning, and cost-aware stopping; and (III)~cascade deployment over the remaining unlabeled workload.}
    \label{fig:actis-schematic}
\end{figure*}
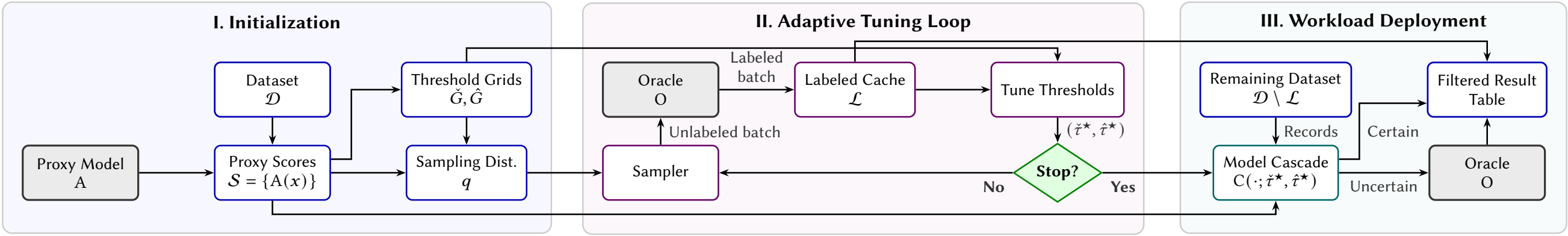

For observations bounded in the unit interval $Z_i \in [0, 1]$, \citet{waudby2024estimating} construct $M_t$ via a betting game. 
Starting from capital $M_0 = 1$, wealth accumulates multiplicatively $M_t = \prod_{i=1}^t \left(1 + \lambda_i (Z_i - m) \right)$, where $m$ is the target margin, $\lambda_i \ge 0$ is a betting fraction determined by past observations $Z_1, \dots, Z_{i-1}$:
\begin{equation}
    \lambda_i = \min\left( \sqrt{\frac{2\log(1/\alpha)}{\tilde{\sigma}_{i-1}^2 \, i \log(1 + i)}}, \; \frac{c}{m} \right).
    \label{eqn:betting-fraction}
\end{equation}
Here $c \in (0, 1)$ ensures that the wealth multiplier remains strictly positive, and $\tilde{\sigma}_t^2$ is an online regularized sample variance estimator:
\begin{equation*}
    \tilde{\sigma}_t^2 = \frac{\frac{1}{4} + \sum_{i=1}^t (Z_i - \tilde{\mu}_i)^2}{t + 1} \quad \text{with} \quad \tilde{\mu}_t = \frac{\frac{1}{2} + \sum_{i=1}^t Z_i}{t + 1}.
\end{equation*}
This betting supermartingale guarantees non-asymptotic Type~I error control $\le \alpha$ for any sample size $t \ge 1$.\footnote{\citet{waudby2024estimating} combine opposing bets with a hedging weight $\theta \in (0, 1)$ to construct two-sided confidence sequences. 
Since our focus is on one-sided testing of $H_0: \mu < m$, we allocate full capital to the positive direction ($\theta = 1$).}

While betting martingales naturally accommodate observations bounded in $[0, 1]$, applying them to importance-weighted random variables requires rescaling the observations onto the unit interval using the maximum possible importance weight. 
Since importance weights must be reasonably large to improve sampling efficiency, the rescaled increments are compressed toward zero, forcing the bet size $\lambda_i$ in~\cref{eqn:betting-fraction} to become very small. 
Consequently, the betting martingale suffers from sluggish wealth accumulation and poor sample efficiency under importance sampling.

\subsubsection{Asymptotic Testing via Gaussian Mixtures} \label{sec:asymptotic-test}
To achieve high sample efficiency under importance-weighted observations without suffering from conservative bounds, we adopt the asymptotic anytime-valid testing framework of \citet{waudby2024time}.

Adapting the conjugate Gaussian mixture processes of \citet{robbins1970statistical} and \citet{howard2021time} to one-sided hypothesis testing, we track the cumulative excess drift $S_t = \sum_{i=1}^t (Z_i - m)$ and the empirical quadratic variation $V_t = \sum_{i=1}^t (Z_i - \bar{Z}_t)^2$, where $\bar{Z}_t = \frac{1}{t} \sum_{i=1}^t Z_i$. 
The mixture test process $(M_t)_{t=0}^\infty$ is defined with $M_0 = 1$ and
\begin{equation}
    M_t = \begin{cases}
        \frac{1}{\sqrt{1 + V_t / v}} \exp\left( \frac{S_t^2}{2(v + V_t)} \right), & \text{if } S_t > 0, \\
        0, & \text{if } S_t \le 0,
    \end{cases}
    \label{eqn:gauss-mix-supermartingale}
\end{equation}
where $v > 0$ is a prior variance parameter.\footnote{In sequential testing, $v$ is typically tuned to the expected cumulative process variance over a planned sample horizon $n_0$ (i.e., $v \approx n_0 \sigma^2$) to optimize statistical power.}
Standardizing increments using the empirical variance $V_t$ rather than a known Gaussian variance or worst-case bound means that $M_t$ is not in general a test supermartingale. 
Instead, its validity is established asymptotically via time-uniform central limit theorems~\citep{waudby2024time}:
\begin{equation}
    \limsup_{m_0 \to \infty} \Pr_{H_0}\left(\exists t \ge m_0 \colon M_t \ge \frac{1}{\alpha}\right) \le \alpha,
    \label{eqn:asym-error-control}
\end{equation}
where $m_0$ denotes the starting sample size for testing. 

While Gaussian mixture processes provide the variance-adaptive power needed for importance sampling, they rely on asymptotic approximations. 
ACTIS therefore pairs them with finite-sample runtime safeguards (\cref{sec:asymptotic-safeguards}), to balance statistical soundness and cost efficiency, while also supporting betting martingales when strict finite-sample guarantees are prioritized.

\section{ACTIS Framework} \label{sec:method}

This section presents ACTIS---\textbf{A}daptive \textbf{C}ascade \textbf{T}uning via \linebreak[4] \textbf{I}mportance \textbf{S}ampling---our framework for tuning two-threshold model cascades under statistical accuracy guarantees. 
As illustrated in \cref{fig:actis-schematic}, ACTIS executes a semantic filter over a workload $\D$ in three stages:

\myparagraph{Stage I: Offline Initialization.}
Before querying the oracle, the proxy model $\proxy$ evaluates all $N$ records in $\D$ to produce proxy scores $\mathcal{S} = \{\proxy(x) : x \in \D\}$. 
From $\mathcal{S}$, ACTIS constructs discrete lower and upper candidate threshold grids $\lo{\grid}$ and $\up{\grid}$ (\cref{sec:grids}), along with an instrumental sampling distribution $q$ designed to minimize the asymptotic variance of the true positive rate, false positive rate and prevalence estimators across the grids (\cref{sec:instrumental-design}).

\myparagraph{Stage II: Adaptive Tuning Loop (\cref{alg:tuning}).}
Rather than committing to a fixed pilot sample size upfront, ACTIS acquires oracle labels incrementally in batches of size $B$ sampled from $\D$ according to $q$ (or uniformly) and appends them to a labeled cache $\mathcal{L}$. 
After each batch, ACTIS updates the cascade thresholds $(\lo{\tau}^\star, \up{\tau}^\star) \in \lo{\grid} \times (\up{\grid} \cup \{\infty\})$ to be as aggressive as possible while certifying the joint precision and recall targets at family-wise error rate $\delta$. 
Individual candidate thresholds are verified using anytime-valid tests (\cref{sec:single-threshold-test})---instantiated via non-asymptotic betting martingales or asymptotic Gaussian mixtures with finite-sample safeguards (\cref{sec:asymptotic-safeguards})---while directional monotonicity decouples the 2D grid search into two 1D fixed-sequence scans without conservative Bonferroni penalties (\cref{sec:tuning-algorithm}). 
After updating $(\lo{\tau}^\star, \up{\tau}^\star)$, a Value-of-Information stopping rule (\cref{sec:stopping-condition}) checks whether the expected reduction in oracle calls over the remaining dataset $\D \setminus \mathcal{L}$ justifies the cost of labeling another batch; if not, the tuning loop halts.

\myparagraph{Stage III: Workload Deployment.}
Upon termination of the tuning loop, the tuned cascade $\cascade(\cdot; \lo{\tau}^\star, \up{\tau}^\star)$ is deployed over the remaining unlabeled tuples $\D \setminus \mathcal{L}$. 
Tuples with proxy scores in $[\lo{\tau}^\star, \up{\tau}^\star)$ are escalated to the oracle $\oracle$, while confident tuples ($\proxy(x) < \lo{\tau}^\star$ or $\proxy(x) \ge \up{\tau}^\star$) are filtered directly using the proxy predictions. 
Combined with the already-labeled cache $\mathcal{L}$, this produces the final filtered result table.

The remainder of this section details each component of ACTIS. All proofs are deferred to \ifarxiv\cref{app:proofs}\else\citet[App.~A]{actis_full}\fi.

\subsection{Candidate Threshold Grids} \label{sec:grids}

Tuning a cascade over thresholds $\lo{\tau} \in [0, 1]$ and $\up{\tau} \in [\lo{\tau}, 1] \cup \{\infty\}$ poses a severe multiple-testing challenge: a continuous search requires either loose uniform concentration bounds or massive samples to prevent overfitting. 
ACTIS restricts tuning to discrete grids of candidate thresholds: $\lo{\grid} = (\lo{\tau}_{1} < \ldots < \lo{\tau}_{\lo{m}})$ for the lower candidates and $\up{\grid} = (\up{\tau}_{1} < \cdots < \up{\tau}_{\up{m}})$ for the upper candidates.
These grids are constructed from the empirical quantiles of the proxy scores.
Specifically, for a target grid size $M$ with $K = \min(M, N)$, we evaluate quantile points $p_i = 1 - (1 - \frac{i - 1}{K - 1})^\kappa$ for $i \in \{1, \dots, K\}$, and select scores at ranks $k_i = 1 + \lfloor (N - 1) p_i \rfloor$ as the threshold grid.
The tail exponent $\kappa \ge 1$ controls candidate concentration.

\myparagraph{Lower Candidate Grid ($\lo{\grid}$).}
The lower threshold $\lo{\tau}$ screens negatives to satisfy the recall target $\gamma_R$. 
Errors (false negatives) are typically distributed across the broad low- to mid-range of proxy scores. 
We therefore construct $\lo{\grid}$ using a uniform quantile schedule ($\kappa = 1$) placing candidate thresholds at equal increments of empirical probability mass to ensure balanced resolution across the bulk of the score distribution.

\myparagraph{Upper Candidate Grid ($\up{\grid}$).}
The upper threshold $\up{\tau}$ governs positive predictions to satisfy the precision target $\gamma_P$. 
Under realistic targets ($\gamma_P \to 1$), cascade precision is highly sensitive to small threshold shifts in the upper tail near~1. 
A uniform quantile schedule is ill-suited here, as it allocates most candidate thresholds to uninformative low scores and leaves negligible resolution in the upper tail. 
We therefore construct $\up{\grid}$ using a power-law schedule with tail exponent $\kappa = \max\left(1, \, \log N / \log K\right)$, which concentrates candidate thresholds in the upper tail and adapts candidate resolution to workload size $N$ without inflating the grid size $\up{m}$.

\myparagraph{Anchor Thresholds.}
To guarantee safe fallbacks for any accuracy target, we pair the lower anchor $\lo{\tau}_{1} = \min_{x \in \D} \proxy(x) \in \lo{\grid}$ with an upper sentinel $\up{\tau}_{\up{m}+1} \coloneqq \infty \notin \up{\grid}$ (strictly exceeding $\max_{x \in \D} \proxy(x)$).
Since the cascade filters on $\proxy(x) < \lo{\tau}$ and $\proxy(x) \ge \up{\tau}$, choosing $\lo{\tau}_{1}$ prevents false negatives ($\Recall(\lo{\tau}_{1}) = 1$) while choosing $\up{\tau}_{\up{m}+1}$ prevents false positives ($\Precision(\lo{\tau}, \up{\tau}_{\up{m}+1}) = 1$).

\subsection{Verifying Individual Cascade Constraints} \label{sec:single-threshold-test}

Before addressing the search over candidate threshold grids, we first consider testing whether individual precision\slash recall constraints are satisfied at fixed candidate thresholds $(\lo{\tau}, \up{\tau}) \in \lo{\grid} \times \up{\grid}$. 
Suppose that after drawing $n$ samples, the tuner has collected an oracle-labeled set $\mathcal{L} = \{(x_i, y_i)\}_{i=1}^n$, where each record $x_i$ was sampled from instrumental distribution $q$ over $\D$ with associated importance weight $w(x_i) = p(x_i) / q(x_i)$. 
Under uniform sampling, $q$ is the nominal distribution $p$, yielding unit weights $w(x_i) \equiv 1$.

The precision\slash recall constraints in~\cref{eqn:opt-problem} are expressed as nonlinear ratios of expectations. 
However, the anytime-valid testing frameworks introduced in \cref{sec:anytime-tests} evaluate whether a population mean (i.e., a single expectation) exceeds a target margin. 
To apply these tests, we reformulate both the recall and precision constraints into equivalent linear expectations, derive their per-sample increments, and define testing subroutines for each constraint.

\subsubsection{Verifying the Recall Constraint} \label{sec:recall-verification}
Recall depends only on the lower threshold $\lo{\tau}$. 
Multiplying~\cref{eqn:recall} by the prevalence $\Prevalence$, the constraint $\Recall(\lo{\tau}) \ge \gamma_R$ is equivalent to a non-negative \emph{recall margin}:
\begin{equation}
    \Delta_R(\lo{\tau}) \coloneqq \E_{X \sim q, Y \sim \oracle(X)}\bigl[f_R(X, Y; \lo{\tau})\bigr] \ge 0,
    \label{eqn:recall-margin}
\end{equation}
where the per-sample increment is defined as:
\begin{equation}
    f_R(X, Y; \lo{\tau}) \coloneqq w(X) Y \bigl(\indicator{\proxy(X) \ge \lo{\tau}} - \gamma_R\bigr).
    \label{eqn:recall-increment}
\end{equation}
To certify that candidate threshold $\lo{\tau}$ meets the recall target, we test the one-sided null hypothesis that the recall margin is negative, $H_{0, R}(\lo{\tau}) \colon \Delta_R(\lo{\tau}) < 0$ versus $H_{1, R}(\lo{\tau}) \colon \Delta_R(\lo{\tau}) \ge 0$.
Rejecting $H_{0, R}(\lo{\tau})$ certifies $\Recall(\lo{\tau}) \ge \gamma_R$. 
For the asymptotic Gaussian mixture test this hypothesis is evaluated directly against the null margin $m_R = 0$.

\subsubsection{Verifying the Precision Constraint} \label{sec:precision-verification}
Precision depends on both $\lo{\tau}$ (which governs true positives) and $\up{\tau}$ (which governs false positives). 
Rearranging~\cref{eqn:precision}, the constraint $\Precision(\lo{\tau}, \up{\tau}) \ge \gamma_P$ is equivalent to a non-negative \emph{precision margin}:
\begin{equation}
    \Delta_P(\lo{\tau}, \up{\tau}) \coloneqq \E_{X \sim q, Y \sim \oracle(X)}\bigl[f_P(X, Y; \lo{\tau}, \up{\tau})\bigr] \ge 0,
    \label{eqn:precision-margin}
\end{equation}
where the per-sample increment is defined as:
\begin{equation}
    f_P(X, Y; \lo{\tau}, \up{\tau}) \coloneqq w(X) \big( (1 - \gamma_P) Y \indicator{\proxy(X) \ge \lo{\tau}} 
        - \gamma_P (1 - Y) \indicator{\proxy(X) \ge \up{\tau}} \big).
    \label{eqn:precision-increment}
\end{equation}
To certify that threshold pair $(\lo{\tau}, \up{\tau})$ meets the precision target, we test the one-sided null hypothesis, $H_{0, P}(\lo{\tau}, \up{\tau}) \colon \Delta_P(\lo{\tau}, \up{\tau}) < 0$ versus $H_{1, P}(\lo{\tau}, \up{\tau}) \colon \Delta_P(\lo{\tau}, \up{\tau}) \ge 0$.
Rejecting $H_{0, P}(\lo{\tau}, \up{\tau})$ certifies that $\Precision(\lo{\tau}, \up{\tau}) \ge \gamma_P$. 
As with recall, the asymptotic test evaluates this hypothesis directly against the margin $m_P = 0$.

\subsubsection{Testing Subroutine Interfaces} \label{sec:subroutine-interfaces}
We encapsulate constraint verification for recall and precision in two boolean subroutines:
\begin{itemize}[leftmargin=1.5em]
    \item $\Call{RecallMet}{\gamma_R, \mathcal{L}, \lo{\tau}, \alpha}$: Tests the null $H_{0, R}(\lo{\tau})$ at significance level $\alpha$ using labeled sample $\mathcal{L}$, returning \texttt{True} if the null is rejected and \texttt{False} otherwise.
    \item $\Call{PrecisionMet}{\gamma_P, \mathcal{L}, \lo{\tau}, \up{\tau}, \alpha}$: Tests the null $H_{0, P}(\lo{\tau}, \up{\tau})$ at significance level $\alpha$ using labeled sample $\mathcal{L}$, returning \texttt{True} if the null is rejected and \texttt{False} otherwise.
\end{itemize}
For the non-asymptotic betting test (\cref{sec:non-asymptotic-test}), raw increments are mapped to $[0, 1]$ via an affine transformation \ifarxiv(\cref{app:rescaling})\else\citep[App.~A.1]{actis_full}\fi. 
The subroutine evaluates wealth accumulation using default betting hyperparameters \ifarxiv(\cref{app:test-hyperparameters})\else\citep[App.~A.2]{actis_full}\fi, returning \texttt{True} once the accumulated wealth satisfies $M_t \ge 1/\alpha$. 
For the asymptotic Gaussian mixture test \ifarxiv(\cref{sec:asymptotic-test}), increments are evaluated directly against margin $m = 0$. 
Its prior variance parameter is set to an estimated horizon variance $v^{\text{est}}$ (\cref{app:test-hyperparameters}). 
Candidate thresholds must additionally satisfy the runtime safeguards detailed next\else, increments are tested against $m = 0$ with prior variance $v^{\text{est}}$ estimated offline from proxy scores at the null boundary over horizon $n_0 = \min (1000, \max(50, N/10))$ \citep[App.~A.2]{actis_full}, subject to the runtime safeguards below\fi.

\subsection{Safeguards for Asymptotic Testing} \label{sec:asymptotic-safeguards}

The Gaussian mixture test introduced in \cref{sec:asymptotic-test} provides anytime-valid error control in the asymptotic limit ($m_0 \to \infty$) via a time-uniform central limit theorem (CLT)~\citep{waudby2024time}. 
However, deploying asymptotic tests on finite, highly skewed workloads---particularly under strict precision\slash recall targets and importance sampling---can expose failure modes:
\begin{itemize}[leftmargin=1.5em]
    \item \emph{Premature Single-Sample Rejections}: In early samples ($t \approx 1$), empirical variance has not yet accumulated ($V_t = 0$). If the prior variance parameter $v$ is too small, an isolated positive increment can trigger an immediate boundary crossing ($M_1 \ge 1/\alpha$) on the first sample.
    \item \emph{Early Success Streaks}: Under strict targets, constraint violations (false negatives or false positives) are rare under the boundary null. In small samples containing no observed violations, early runs of consecutive successes generate positive cumulative drift with zero empirical variance ($V_t = 0$). 
\end{itemize}

To address these vulnerabilities without sacrificing the sample efficiency of asymptotic testing, ACTIS introduces three complementary safeguards.

\subsubsection{Safeguard 1: Single-Step Safety Floor}
To ensure that a single sample cannot cause the wealth $M_t$ to cross the rejection boundary ($1/\alpha$), we clip the prior variance parameter $v$ from below:
\begin{equation}
    v \gets \max\bigl(v, \, v^{\text{floor}}\bigr).
    \label{eqn:v-final}
\end{equation}
Let $b_R(\lo{\tau}) \coloneqq (1 - \gamma_R) \wmax^{\ge}(\lo{\tau})$ and $b_P(\lo{\tau}) \coloneqq (1 - \gamma_P) \wmax^{\ge}(\lo{\tau})$ denote the maximum possible recall and precision increments, where \ifarxiv$\wmax^{\ge}(\cdot)$ is defined in \cref{eqn:importance-weight-bounds}, within \cref{app:rescaling}\else$\wmax^{\ge}(\tau) = \max_{x \in \D \colon \proxy(x) \ge \tau} w(x)$\fi.
For significance levels $\delta_R = \delta / 2$ and $\alpha_P = \delta / (2 \lo{m})$, the safety floor is:
\begin{equation}
    v_R^{\text{floor}}(\lo{\tau}) \coloneqq \frac{2 b_R^2(\lo{\tau})}{\log(1/\delta_R)}, \qquad
    v_P^{\text{floor}}(\lo{\tau}) \coloneqq \frac{2 b_P^2(\lo{\tau})}{\log(1/\alpha_P)}. 
    \label{eqn:v-min}
\end{equation}
As proven in \ifarxiv\cref{app:safeguards}\else~the full paper \citep[App.~A.3]{actis_full}\fi, setting $v \ge v^{\text{floor}}$ guarantees that for any first observation ($V_1 = 0$), accumulated wealth satisfies $M_1 < 1/\alpha$.

\subsubsection{Safeguard 2: Variance Accumulation Check}
The variance-stabilization condition underpinning the time-uniform CLT~\citep{waudby2024time} requires the empirical quadratic variation $V_t$ to grow sufficiently large relative to the prior variance $v$. 
When $V_t$ is negligible, the Gaussian mixture is dominated by the prior rather than empirical observations. 
Before rejecting $H_{0,R}$ or $H_{0,P}$, ACTIS requires the corresponding empirical quadratic variation to reach at least $\rho = 5\%$ of the estimated horizon variance:
$V_t / v^{\text{est}} \ge \rho$.

\subsubsection{Safeguard 3: Binomial Safety Check}
To guard against premature false rejections on early success streaks, ACTIS requires the observed failure count to also pass an exact one-sided binomial test before rejecting $H_{0,R}$ or $H_{0,P}$.
For each test, active records in $\mathcal{L}$ are partitioned into successes and failures:
\begin{itemize}[leftmargin=1.5em]
    \item \emph{Recall at $\lo{\tau}$}: Active records are ground-truth positives ($Y = 1$). A success is a true positive ($\proxy(X) \ge \lo{\tau}$), while a failure is a false negative ($\proxy(X) < \lo{\tau}$).
    \item \emph{Precision at $(\lo{\tau}, \up{\tau})$}: Active records are cascade positive predictions. A success is a true positive ($\proxy(X) \ge \lo{\tau}, Y = 1$), while a failure is a false positive ($\proxy(X) \ge \up{\tau}, Y = 0$).
\end{itemize}
Under uniform random sampling, the conditional failure probability of an active record drawn under the boundary null is $p_0 = 1 - \gamma$ (where $\gamma \in \{\gamma_R, \gamma_P\}$). 
Under importance sampling, the instrumental distribution $q$ allocates non-uniform weight across proxy score regions. 
As derived in \ifarxiv\cref{app:safeguards}\else the full paper \citep[App.~A.3]{actis_full}\fi, the probability of drawing a failure under $q$ given that an active record is sampled shifts to:
\begin{equation*}
    p_0 \coloneqq \frac{(1 - \gamma) \eta}{\gamma + (1 - \gamma) \eta},
\end{equation*}
where $\eta \coloneqq \bar{q}_F / \bar{q}_S$ denotes the ratio of average instrumental densities across the failure and success regions. 

Among the observed batches, suppose there are $n_{\text{a}}$ active records and $n_{\text{f}}$ failures. Safeguard~3 passes if and only if $F(n_{\text{f}}; n_{\text{a}}, p_0) \le \alpha$, where $\alpha = \delta_R$ for recall and $\alpha = \alpha_P$ for precision, and $F(x; n, p)$ denotes the cumulative distribution function of a binomial random variable with $n$ trials and per-trial failure probability $p$.

\subsubsection{Integration into Test Subroutines}
When evaluated via asymptotic testing, $\Call{RecallMet}{}$ and $\Call{PrecisionMet}{}$ return \texttt{True} if and only if the mixture test process crosses the nominal rejection threshold ($M_t \ge 1/\alpha$) and both Safeguards~2 and~3 are satisfied. 

\subsection{Two-Stage Tuning via Fixed-Sequence Testing} \label{sec:tuning-algorithm}

Given candidate threshold grids $\lo{\grid}$ and $\up{\grid}$, the goal of cascade tuning is to find a threshold pair $(\lo{\tau}^\star, \up{\tau}^\star) \in \lo{\grid} \times (\up{\grid} \cup \{\infty\})$ that minimizes the oracle call rate while guaranteeing the recall and precision targets are met with family-wise error rate (FWER) at most $\delta$.

A naive approach would evaluate each threshold pair on the grid using the constraint verification subroutines $\Call{RecallMet}{}$ and $\Call{PrecisionMet}{}$ introduced in \cref{sec:subroutine-interfaces}. 
However, testing all $\lo{m} \cdot \up{m}$ pairs without correcting for multiple hypothesis testing risks inflating the Type~I error. 
Conversely, applying a Bonferroni correction over the grid would test each hypothesis at significance level $\delta / (\lo{m} \cdot \up{m})$. 
In practical configurations with $\lo{m} = 50$ and $\up{m} = 500$, dividing $\delta$ by $25{,}000$ would be very conservative, requiring an impractically large sample $\mathcal{L}$ to reject any null hypothesis.

\subsubsection{Fixed-Sequence Testing}
To avoid the power loss of a 2D~Bonferroni correction, we leverage the \emph{fixed-sequence testing method}~\citep{maurer1995multiple,westfall2001optimally}. 
Under this method, a family of null hypotheses $H_1, \dots, H_M$ is evaluated in a pre-specified sequential order, terminating at the first non-rejection. 
Each individual hypothesis is tested at the full nominal level $\alpha$, yet the family-wise error rate of the test remains bounded by $\alpha$ with no multiplicative penalty.

\subsubsection{Monotonicity and Power}
While fixed-sequence testing guarantees FWER control for any pre-specified sequence, the procedure halts at the first non-rejection. Consequently, an arbitrary hypothesis ordering risks premature termination on an early true null, missing valid candidate thresholds located further down the sequence.
Fortunately, the precision and recall constraints satisfy a monotonicity property that aligns with sequential testing.

\begin{proposition}[Monotonicity of Cascade Constraints] \label[proposition]{prop:monotonicity} 
    For any precision\slash recall targets $\gamma_P, \gamma_R \in (0, 1)$:
    \begin{romanenum} 
        \item The recall margin $\Delta_R(\lo{\tau})$ is monotonically non-increasing in $\lo{\tau}$.
        \item The precision margin $\Delta_P(\lo{\tau}, \up{\tau})$ is monotonically non-decreasing in $\up{\tau}$.
    \end{romanenum}
\end{proposition}

By \cref{prop:monotonicity}, cascade constraint violations are nested. Scanning candidate thresholds monotonically (ascending for $\lo{\tau}$ and descending for $\up{\tau}$) ensures that all true null hypotheses form a contiguous suffix at the end of the sequence, so halting at the first non-rejection incurs no loss of statistical power.

\subsubsection{Two-Stage Tuning Procedure}
Although cascade tuning involves searching over the 2D~grid $\lo{\grid} \times \up{\grid}$, the two constraints exhibit an asymmetry: the recall margin $\Delta_R(\lo{\tau})$ depends exclusively on the lower threshold, whereas the precision margin $\Delta_P(\lo{\tau}, \up{\tau})$ depends on both thresholds. 
This asymmetry allows us to decouple the 2D search into two 1D stages: we first optimize the lower threshold $\lo{\tau}$ against the recall constraint alone, and subsequently optimize the upper threshold $\up{\tau}$ conditioned on the selected lower threshold.

To ensure that both constraints hold jointly with FWER at most $\delta$, we split the error budget equally between the two stages via a union bound, allocating $\delta_R = \delta / 2$ for recall and $\delta_P = \delta / 2$ for precision. 
After each batch is added to $\mathcal{L}$, threshold updates proceed in two stages (lines~\ref{alg:tuning:start-tune}--\ref{alg:tuning:end-tune} of \cref{alg:tuning}):
\begin{enumerate}[leftmargin=1.5em]
    \item \emph{Forward Scan for $\lo{\tau}$.} 
    Starting from safe anchor threshold $\lo{\tau}_{1}$, we evaluate candidate lower thresholds in ascending order at level $\delta_R$ until the first non-rejection of the recall null $H_{0, R}$, selecting the largest passing threshold as $\lo{\tau}^\star$. 
    \item \emph{Backward Scan for $\up{\tau}$.} 
    Conditioning on the data-dependent $\lo{\tau}^\star$, we evaluate candidate upper thresholds in descending order starting from anchor threshold $\up{\tau}_{\up{m} + 1} = \infty$ down to $\lo{\tau}^\star$ at Bonferroni-adjusted level $\alpha_P = \delta_P / \lo{m}$, stopping at the first non-rejection to select the smallest passing threshold $\up{\tau}^\star$. 
\end{enumerate}

\begin{algorithm}[t]
\caption{ACTIS Adaptive Cascade Tuning}\label{alg:tuning}
\begin{algorithmic}[1]
\Require Dataset $\D$, proxy scores $\mathcal{S}$, grids $\lo{\grid}, \up{\grid}$, sampling distribution $q$, batch size $B$, accuracy targets $\gamma_P, \gamma_R$, error budget $\delta$.
\Ensure Tuned thresholds $(\lo{\tau}^\star, \up{\tau}^\star)$, labeled cache $\mathcal{L}$
\State $\mathcal{L} \gets \emptyset$, \quad $\delta_R \gets \delta / 2$, \quad $\delta_P \gets \delta / 2$, \quad $\alpha_P \gets \delta_P / \lo{m}$
\Repeat
    \State Sample batch $\mathcal{B}$ from $\D$ of size $B$ via $q$ (or uniformly)
    \State Query oracle on $\mathcal{B}$ and add to labeled cache $\mathcal{L}$
    \State $\lo{k} \gets 1$ \Comment{Stage 1: Forward scan for $\lo{\tau}^\star$} \label{alg:tuning:start-tune}
    \While{$\lo{k} < \lo{m}$ \textbf{and} $\Call{RecallMet}{\gamma_R, \mathcal{L}, \lo{\tau}_{\lo{k} + 1}, \delta_R}$}
        \State $\lo{k} \gets \lo{k} + 1$
    \EndWhile
    \State $\lo{\tau}^\star \gets \lo{\tau}_{\lo{k}}$
    \State $\up{k} \gets \up{m} + 1$ \Comment{Stage 2: Backward scan for $\up{\tau}^\star$}
    \While{$\up{k} > 1$ \textbf{and} $\up{\tau}_{\up{k} - 1} \ge \lo{\tau}^\star$ \textbf{and} $\Call{PrecisionMet}{\gamma_P, \mathcal{L}, \lo{\tau}^\star, \up{\tau}_{\up{k} - 1}, \alpha_P}$}
        \State $\up{k} \gets \up{k} - 1$
    \EndWhile
    \State $\up{\tau}^\star \gets \up{\tau}_{\up{k}}$ \label{alg:tuning:end-tune}
\Until{$\Call{ShouldStop}{N, \mathcal{L}, \lo{\tau}^\star, \up{\tau}^\star, B}$ (see \cref{sec:stopping-condition})}
\State \Return $(\lo{\tau}^\star, \up{\tau}^\star), \mathcal{L}$
\end{algorithmic}
\end{algorithm}

\begin{proposition}[Family-Wise Error Rate Control] \label[proposition]{prop:fwer-control}
    Given anytime-valid test subroutines \Call{RecallMet}{} and \Call{PrecisionMet}{}, \cref{alg:tuning} satisfies the joint precision\slash recall guarantee in~\cref{eqn:opt-problem} with family-wise error rate (FWER) at most $\delta$.
\end{proposition}

By decoupling tuning in this way, fixed-sequence testing avoids any multiplicative penalty over the upper grid size $\up{m}$. 
This allows ACTIS to use dense grids for the upper threshold (e.g., $\up{m} = 500$) to maximize precision without sacrificing statistical power. 

\subsection{Adaptive Stopping via Marginal Value of Information} \label{sec:stopping-condition}

In an adaptive deployment, oracle labels are acquired sequentially in batches of size $B$ sampled from the workload $\D$. 
Let $n = \abs{\mathcal{L}}$ denote the number of labeled records collected so far across all batches. 
At each step, the tuner must decide whether to spend oracle calls labeling another batch of size
$B$ to tighten the thresholds, or stop tuning and deploy the cascade over the remaining workload.

\myparagraph{Anytime Validity and Arbitrary Stopping.}
A fundamental advantage of the anytime-valid test processes introduced in \cref{sec:anytime-tests} is that their statistical error guarantees hold under optional stopping. 
By Ville's inequality and Doob's optional stopping theorem~\citep{howard2021time}, the Type~I error bound $\delta$ holds at \emph{any arbitrary, data-dependent stopping time}. 
Consequently, the tuner is free to stop whenever continuing is no longer cost-effective, without inflating the FWER or invalidating the cascade guarantees.

\myparagraph{Optimistic Thresholds.}
To quantify how much the cascade could potentially improve with additional data, we invert the anytime tests to construct \emph{optimistic thresholds}.
Recall that \cref{alg:tuning} selects conservative thresholds by rejecting the null hypothesis that the quality margin is negative ($H_0 \colon \Delta < 0$), certifying that the true margin is non-negative (a lower confidence bound on quality). 

Conversely, we define a candidate threshold as \emph{plausible} if the data cannot reject the reverse null hypothesis that the quality margin is non-negative ($H_0 \colon \Delta \ge 0$). 
By testing this reverse hypothesis, we identify the most aggressive candidate thresholds that cannot yet be ruled out as invalid within the anytime confidence bounds. 
Specifically, the \emph{optimistic lower threshold} $\lo{\tau}^\text{opt}$ is the largest candidate in $\lo{\grid}$ where the recall target remains plausible, while the \emph{optimistic upper threshold} $\up{\tau}^\text{opt}$ is the smallest candidate in $\up{\grid} \cup \{\infty\}$ (with $\up{\tau}^\text{opt} \ge \lo{\tau}^\text{opt}$) where the precision target remains plausible. 
These optimistic thresholds are purely an internal heuristic for the stopping rule; they are not deployed or used to accept candidates, and therefore do not consume any FWER error budget.

\myparagraph{Marginal Value of Information.}
Let $r^\star = r(\lo{\tau}^\star, \up{\tau}^\star)$ and $r^\text{opt} = r(\lo{\tau}^\text{opt}, \up{\tau}^\text{opt})$ denote the oracle call rates under the current conservative thresholds and optimistic thresholds, respectively.
Since the plausible region contains the true optimal thresholds with high probability, the difference $\Delta r \coloneqq \max \left(0, \, r^\star - r^\text{opt}\right)$ represents the \emph{uncertainty gap}~\citep{kalyanakrishnan2012pac}---the maximum reduction in oracle call rate that could be gained by collecting infinite additional samples.

Following the principle of the \emph{Value of Information}~\citep{howard1966information}, we compare the expected marginal savings from acquiring the next batch against its marginal acquisition cost:
\begin{itemize}[leftmargin=1.5em]
\item \emph{Expected Marginal Savings}: Confidence sequence widths contract at rate $O(1/\sqrt{n})$. Adding a batch of size $B$ to an existing sample of size $n$ contracts the confidence bound width by a factor of $1 - \sqrt{n / (n + B)} \approx \frac{B}{2n}$. 
Across the $N - n$ remaining records in the workload $\D$, the expected reduction in oracle invocations is
$\operatorname{Savings}(B) \coloneqq \left( \frac{B}{2 \max(1, n)} \right) \cdot \Delta r \cdot (N - n)$. 
\item \emph{Marginal Acquisition Cost}: If tuning were terminated immediately and the cascade deployed, only a fraction $r^\star$ of the incoming batch would be escalated to the oracle. 
On the other hand, continuing tuning requires invoking the oracle on an additional $B$ records, incurring a net marginal cost
$\operatorname{Cost}(B) \coloneqq B \cdot (1 - r^\star)$.
\end{itemize}

\myparagraph{Stopping Condition.}
The $\Call{ShouldStop}{}$ subroutine in \cref{alg:tuning} halts tuning once expected marginal savings no longer exceed marginal acquisition cost, i.e., $\operatorname{Savings}(B) \le \operatorname{Cost}(B)$, or once a maximum tuning budget of $n_{\max}$ samples (default $N/2$) is reached.
To prevent premature stopping under severe class imbalance---where initial batches may contain zero positives---early stopping is deferred until at least $n_\text{min}$ positive instances have been observed \ifarxiv(\cref{app:stopping-details})\else\citep[App.~A.6]{actis_full}\fi.

\subsection{Variance-Minimizing Instrumental Distribution} \label{sec:instrumental-design}

In each step of the adaptive tuning loop, ACTIS acquires a batch of $B$ labeled records. 
Under uniform sampling, records are drawn with equal probability $p(x) = 1/N$. 
When target positives are rare (e.g., prevalence $\Prevalence \le 0.05$), uniform sampling spends the vast majority of its oracle budget acquiring uninformative true negatives, starving the tuner of positive examples and leading to wide confidence sequences. 
To accelerate sequential testing, ACTIS uses importance sampling: it draws records from an instrumental distribution $q \in \Delta(\D)$ biased toward informative records, and weights each observed increment by $w(x) = p(x) / q(x)$.

The statistical efficiency of importance sampling hinges on $q$. 
Rather than adopting ad-hoc heuristics, we derive $q$ following the principle of asymptotic variance minimization across the candidate threshold grids $\lo{\grid}$ and $\up{\grid}$.
Let $\lo{m}(s) \coloneqq \sum_{\tau \in \lo{\grid}} \indicator{s \ge \tau}$ and $\up{m}(s) \coloneqq \sum_{\tau \in \up{\grid}} \indicator{s \ge \tau}$ denote the number of candidate lower and upper thresholds that fall at or below proxy score $s$.

\begin{proposition} \label[proposition]{prop:opt-instrumental}
    Under a calibrated proxy model ($\E[Y \mid \proxy(x) = s] = s$), the instrumental distribution minimizing the variance of the true positive rate, false positive rate, and prevalence estimators is
    $q^*(x) \propto \sqrt{c(\proxy(x))}$, where the score sensitivity function $c(s)$ is:
    \begin{equation}
        c(s) = \lambda_{\TPR} \, s \left(\frac{\lo{m}(s)}{\lo{m}}\right) 
             + \lambda_{\FPR} \, (1 - s) \left(\frac{\up{m}(s)}{\up{m}}\right) 
             + \lambda_{\Prevalence} \, s.
        \label{eqn:score-sensitivity}
    \end{equation}
\end{proposition}

The component weights $\lambda_{\TPR} = 1 + (\frac{1 - \gamma_P}{\gamma_P})^2$, $\lambda_{\FPR} = 1$, and $\lambda_{\Prevalence} = (\frac{\gamma_R}{1 - \gamma_R})^2$ standardize the recall and precision constraints by their slack scales $1 - \gamma_R$ and $\gamma_P$; they concentrate sampling priority on discovering true positives for recall ($\lambda_{\TPR} \ge 1$) and detecting false positives for precision ($\lambda_{\FPR} = 1$), while scaling up prevalence discovery for practical recall targets ($\gamma_R \to 1$). 
The joint variance objective, weight derivation, and proof of \cref{prop:opt-instrumental} are deferred to \ifarxiv\cref{app:opt-instrumental}\else the full paper \citep[App.~A.7]{actis_full}\fi.

In practice, proxy models are imperfect: LLM proxies like GPT-5.6-luna often output saturated scores of 0 and 1 (see \ifarxiv\cref{tab:dataset-stats}\else Table~4 of the full paper \citep{actis_full}\fi), and proxies may be miscalibrated across the score domain. 
To defend against this, ACTIS employs two mechanisms:

\myparagraph{Tail Tolerance.}
Since $\up{m}(0) / \up{m} \approx 0$, plugging $s = 0$ into \cref{eqn:score-sensitivity} yields $c(0) \approx 0$, meaning the instrumental distribution $q^*(x)$ assigns negligible weight to 0-scored items. 
This creates a blind spot that ignores potential false negatives. 
To guard against saturated scores without excessive distortion, ACTIS evaluates $q^*(x)$ by replacing raw proxy score $s$ with a clipped proxy score
$\operatorname{clip} (s, \, \nu \hat{\Prevalence}, \, 1 - \nu (1 - \hat{\Prevalence}) )$,
where $\hat{\Prevalence} \coloneqq \frac{1}{N}\sum_{x \in \D}\proxy(x)$ estimates the positive prevalence and $\nu \in (0, 1)$ is the tail tolerance parameter (we set $\nu = 0.1$ by default). 

\myparagraph{Defensive Mixture.}
To prevent importance weights %
from exploding under a poorly-calibrated proxy, ACTIS mixes $q^*$ with the uniform distribution
$q(x) \gets \epsilon p(x) + (1 - \epsilon) q^*(x)$
where $\epsilon \in (0, 1]$ is a defensive mixing parameter (we set $\epsilon = 0.4$ by default).
This guarantees importance weights are bounded by $1 / \epsilon$.

Since both defensive mechanisms depend only on proxy scores and candidate grids, $q(x)$ can be computed prior to cascade tuning without querying the oracle.

\section{Empirical Evaluation} \label{sec:evaluation}

We empirically evaluate ACTIS across 42 benchmarks spanning 23 datasets\slash queries. Our evaluation addresses four research questions:
\begin{enumerate}%
    \item[RQ1.] \emph{Statistical Soundness Audit:} Do ACTIS and existing cascade tuning methods uphold their statistical accuracy guarantees across diverse workloads and error budgets? (\cref{sec:exp-soundness})
    \item[RQ2.] \emph{Cascade Cost-Effectiveness:} How effective is ACTIS at reducing expensive oracle invocations compared to statistically valid baselines? (\cref{sec:exp-efficiency})
    \item[RQ3.] \emph{Ablations and Sensitivity Analysis:} How do the algorithmic components and hyperparameters of ACTIS influence its cost efficiency and tuning latency? (\cref{sec:exp-ablations})
    \item[RQ4.] \emph{Tuning Runtime and Overhead:} What is the computational latency of ACTIS's tuning phase, and is it practical for database query optimization? (\cref{sec:exp-latency})
\end{enumerate}

\subsection{Experimental Setup} \label{sec:exp-setup}

\subsubsection{Workloads and Benchmarks}
We evaluate ACTIS across 23 \emph{workloads} (each defined as a dataset--query pair) spanning computer vision, information extraction, legal contracts, review sentiment, and technical documents~\citep{kang2020approximate,zeighami2025cut,zhang2026scaledoc}. 
As summarized in \cref{tab:domain-summary}, workload sizes range from $N = 1{,}000$ (Court) to $N = 973{,}085$ (Jackson), with positive prevalence $\Prevalence$ (pos.\ rate) spanning three orders of magnitude (from $0.1\%$ to $86.5\%$). 

We define a \emph{benchmark} as a workload paired with a specific proxy model. 
For the 19 text-based workloads derived from \citet{zeighami2025cut} and \citet{zhang2026scaledoc}, we evaluate both cloud-hosted and local proxy models. 
The remaining 4~workloads from \citet{kang2020approximate} were released with precomputed proxy scores and oracle labels without retaining the raw data, contributing one fixed benchmark each. 
In total, this yields 42~benchmarks.
Evaluating these benchmarks across target pairs $(\gamma, \delta)$ produces our full set of experimental scenarios. 
A comprehensive per-benchmark breakdown is provided in \ifarxiv\cref{app:evaluation} (\cref{tab:dataset-stats})\else\citep[Table~4]{actis_full}\fi.

\begin{table}
  \centering
  \caption{Summary of experimental benchmarks. Detailed per-benchmark statistics are reported in \ifarxiv\cref{tab:dataset-stats} in \cref{app:evaluation}\else Table~4 of~\citep{actis_full}\fi.}
  \label{tab:domain-summary}
  \footnotesize
  \setlength{\tabcolsep}{3.5pt}
  \addvspace{-4pt}
  \begin{tabular}{l c c c c}
    \toprule
    \textbf{Benchmark Suite} & \textbf{Benchmarks} & \textbf{Size ($N$)} & \textbf{Pos.\ rate} & \textbf{PR-AUC} \\
    \midrule
    \emph{Pre-LLM Vision \& NLP}~\citep{kang2020approximate} & 4 & 11--973k & 0.1--29\% & 0.79--0.98 \\
    {\scriptsize ImageNet, Jackson, Onto, TACRED} & & & & \\
    \addlinespace[2pt]
    \emph{General Text Corpora}~\citep{zeighami2025cut} & 8 & 1--10k & 5.9--87\% & 0.12--0.97 \\
    {\scriptsize Court, Review, Screenplay, Wiki} & & & & \\
    \addlinespace[2pt]
    \emph{Technical Documents}~\citep{zhang2026scaledoc} & 30 & 10k & 0.6--79\% & 0.004--1.00 \\
    {\scriptsize PubMed, BigPatent, GovReport} & & & & \\
    \midrule
    \textbf{Total\slash Overall} & 42 & 1--973k & 0.1--87\% & 0.004--1.00 \\
    \bottomrule
  \end{tabular}
\end{table}

\subsubsection{Oracle and Proxy Models}
For text benchmarks from \citet{zeighami2025cut} and \citet{zhang2026scaledoc}, we instantiate the oracle using GPT-5.6-terra deployed in Microsoft Foundry. 
We disable reasoning and set the sampling temperature to zero to ensure deterministic oracle ground truth. 
A minimal content filter is applied by Microsoft at the API level; records triggering the filter ($\le 0.4\%$ across all datasets) are excluded. 
For the four vision and NLP benchmarks from \citet{kang2020approximate}, we use the precomputed oracle labels and proxy scores released by the authors.

For the proxy model, we evaluate two deployment regimes:
\begin{itemize}[leftmargin=1.5em]
    \item \emph{Cloud Proxy (GPT-5.6-luna).} A lightweight model from the same family as the oracle (also deployed in Microsoft Foundry) with $10\times$ lower API inference costs. 
    Confidence scores are extracted from output logits for the tokens \texttt{True} and \texttt{False}.
    \item \emph{Local Proxy (Llama-3.2-1B).} A small open-weight language model deployed locally using Ollama on a single NVIDIA A100 GPU (80GB VRAM) supporting batched inference.
\end{itemize}

\subsubsection{Baseline Algorithms}
We compare ACTIS against three state-of-the-art cascade tuning algorithms:

\myparagraph{BARGAIN.} An extension of BARGAIN~\citep{zeighami2025cut} developed to tune two-threshold cascades under joint precision and recall targets. While not present in the BARGAIN paper, its code was released in the authors' open-source repository~\citep{shankar2026bargainpr}. 
As it represents the only prior baseline that is statistically sound, we evaluate it as a primary baseline. 
To enable execution on large workloads, we forked the authors' repository to incorporate NumPy vectorization and configured its batch size to 100.

\myparagraph{LOTUS~\citep{patel2025semantic}.} The cascade tuning algorithm implemented in the LOTUS semantic query engine. We configure it with the default pilot sample size of $10\%$. %
To allow LOTUS to scale to our larger workloads, we forked the authors' repository and re-implemented its \texttt{learn\_cascade\_thresholds} routine using NumPy prefix sums and binary search, reducing its computational complexity from nested iterative loops to $O(N \log N)$.

\myparagraph{ScaleDoc~\citep{zhang2026scaledoc}.} The tuning algorithm from ScaleDoc, evaluated using the authors' published implementation.
Unlike ACTIS, BARGAIN, and LOTUS, which tune cascades to satisfy joint precision and recall targets, ScaleDoc guarantees that the cascade exceeds a user-specified F1 score target $\gamma_{F1}$ with probability $1 - \delta$.

\subsubsection{ACTIS Configuration}
We evaluate three configurations:
\begin{itemize}[leftmargin=1.5em]
    \item \emph{ACTIS} (default): our primary variant, pairing importance sampling with asymptotic Gaussian mixture tests and runtime safeguards ($\epsilon = 0.4$, $\rho = 0.05$, $\nu = 0.1$);
    \item \emph{ACTIS-U}: a non-asymptotic variant combining uniform random sampling with betting supermartingales, guaranteeing finite-sample validity; and
    \item \emph{ACTIS~(w/o safeguards)}: an ablation of the default variant omitting safeguards to isolate their impact on statistical soundness.
\end{itemize}
All configurations use grid sizes $\lo{m} = 50$ and $\up{m} = 500$, batch size $B = 100$, and $n_{\max} = N / 2$
unless otherwise specified.

\subsubsection{Simulation Protocol}
To evaluate both average performance and tail failure behavior, we execute 1,000 independent trials for each experimental scenario. 
In each trial, the tuning algorithm queries a sample of records from $\D$ (one-shot for static baselines, sequentially in batches for adaptive methods) to tune thresholds $(\lo{\tau}^\star, \up{\tau}^\star)$. 
The cascade is then deployed over the remaining workload (not sampled during tuning), reusing oracle labels for records evaluated during tuning. 
We measure the resulting precision, recall, and oracle call rate across the entire workload $\D$ against ground-truth oracle labels. 
All experiments evaluate four accuracy targets $\gamma \in \{0.70, 0.80, 0.90, 0.95\}$ under three nominal error budgets $\delta \in \{0.01, 0.05, 0.10\}$.
For LOTUS, BARGAIN, and ACTIS, $\gamma$ specifies a symmetric precision and recall target ($\gamma_P = \gamma_R = \gamma$). 
Since ScaleDoc natively guarantees only an F1 target, we configure it according to the evaluation goal:
(i)~In RQ1 (auditing), we audit ScaleDoc under its native F1 guarantee by setting $\gamma_{F1} = \gamma$.
(ii)~In RQ2--RQ4 (efficiency), algorithms must achieve the same precision and recall targets. 
We therefore configure ScaleDoc with an F1 target of $\gamma_{F1} = \frac{2\gamma}{1+\gamma}$, which implies $\Precision \ge \gamma$ and $\Recall \ge \gamma$. 

\subsection{RQ1: Auditing Statistical Soundness} \label{sec:exp-soundness}

To assess whether a cascade tuning algorithm satisfies the theoretical guarantee defined in \cref{eqn:opt-problem}, we evaluate its empirical failure rate $\tilde{\delta}$---the fraction of $1{,}000$ simulation trials in which the deployed cascade violates its native accuracy target (precision and recall target for ACTIS, BARGAIN, and LOTUS; F1 target for ScaleDoc). 
As tail probabilities are sensitive to finite-sample variance, we compute a $95\%$ Clopper-Pearson binomial confidence interval $[\tilde{\delta}_\text{LB}, \tilde{\delta}_\text{UB}]$ for $\delta$. 
An experimental scenario is certified \emph{sound} if the upper bound satisfies the budget ($\tilde{\delta}_\text{UB} < \delta$), certified \emph{unsound} if the lower bound exceeds the budget ($\tilde{\delta}_\text{LB} > \delta$), and \emph{inconclusive} if the nominal budget falls within the interval ($\delta \in [\tilde{\delta}_\text{LB}, \tilde{\delta}_\text{UB}]$). 
Inconclusive outcomes indicate that 1,000 trials cannot statistically distinguish $\tilde{\delta}$ from $\delta$, keeping both soundness and unsoundness certifications conservative.

\cref{tab:soundness_summary_joint} summarizes the audit across all 42 benchmarks aggregated over accuracy targets $\gamma \in \{0.70, 0.80, 0.90, 0.95\}$. 
A disaggregated breakdown across each individual $(\gamma, \delta)$ pair is reported in \ifarxiv\cref{app:evaluation} (\cref{tab:soundness_disaggregated_joint})\else\citep[Table~5]{actis_full}\fi.

\begin{table}
\centering
\caption{Statistical soundness across 42 benchmarks by error budget $\delta$, reporting the percentage certified sound, inconclusive, or unsound (highlighted in \badvalue{red}) via 95\% CIs over 1,000 trials.
$\tilde{\delta}_\text{unsound}$ and $\tilde{\delta}_{\max}$ denote the mean empirical failure rate over unsound cases and the maximum across all cases.}
\label{tab:soundness_summary_joint}
\footnotesize
\setlength{\tabcolsep}{3.5pt}
\addvspace{-4pt}
\begin{tabular}{ll rrr rr}
\toprule
& & \multicolumn{3}{c}{\textbf{Percentage Certified}} & & \\
\cmidrule(lr){3-5}
$\delta$ & \textbf{Method} & \textbf{Sound} & \textbf{Inconcl.} & \textbf{Unsound} & $\tilde{\delta}_\text{unsound}$ & $\tilde{\delta}_{\max}$ \\
\midrule
0.01 & ACTIS & 99.4 & 0.6 & 0.0 & \text{---} & 0.004 \\
 & ACTIS (w/o safeguards) & 94.0 & 2.4 & \badvalue{3.6} & 0.113 & 0.217 \\
 & ACTIS-U & 100.0 & 0.0 & 0.0 & \text{---} & 0.001 \\
 & BARGAIN & 91.1 & 8.9 & 0.0 & \text{---} & 0.006 \\
 & ScaleDoc & 44.0 & 3.0 & \badvalue{53.0} & 0.329 & 1.000 \\
 & LOTUS & 15.5 & 3.0 & \badvalue{81.5} & 0.844 & 1.000 \\
\midrule
0.05 & ACTIS & 99.4 & 0.6 & 0.0 & \text{---} & 0.043 \\
 & ACTIS (w/o safeguards) & 95.8 & 0.6 & \badvalue{3.6} & 0.160 & 0.232 \\
 & ACTIS-U & 100.0 & 0.0 & 0.0 & \text{---} & 0.002 \\
 & BARGAIN & 100.0 & 0.0 & 0.0 & \text{---} & 0.030 \\
 & ScaleDoc & 51.8 & 8.3 & \badvalue{39.9} & 0.423 & 1.000 \\
 & LOTUS & 18.5 & 1.2 & \badvalue{80.4} & 0.872 & 1.000 \\
\midrule
0.10 & ACTIS & 100.0 & 0.0 & 0.0 & \text{---} & 0.065 \\
 & ACTIS (w/o safeguards) & 96.4 & 0.6 & \badvalue{3.0} & 0.206 & 0.283 \\
 & ACTIS-U & 100.0 & 0.0 & 0.0 & \text{---} & 0.004 \\
 & BARGAIN & 100.0 & 0.0 & 0.0 & \text{---} & 0.049 \\
 & ScaleDoc & 62.5 & 3.6 & \badvalue{33.9} & 0.481 & 1.000 \\
 & LOTUS & 19.0 & 1.8 & \badvalue{79.2} & 0.893 & 1.000 \\
\bottomrule
\end{tabular}
\end{table}

\begin{figure}
    \centering
    \resizebox{0.85\linewidth}{!}{%
    \input{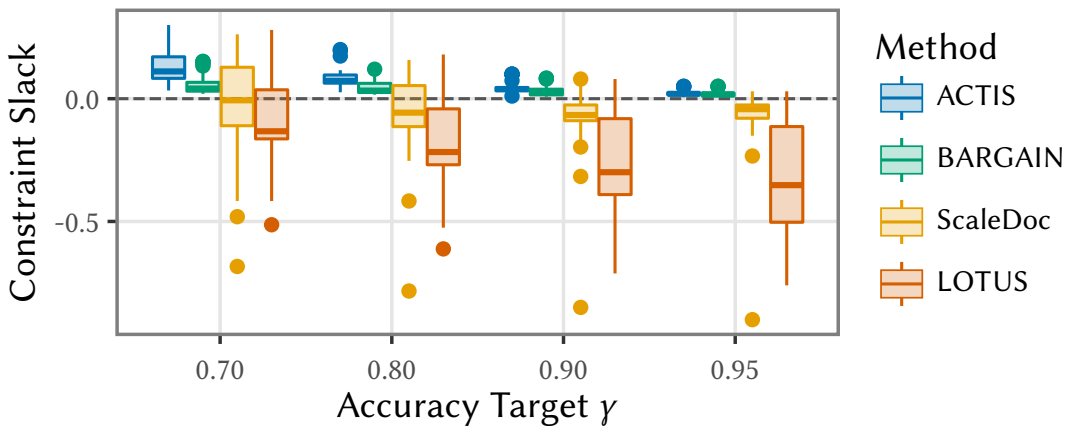}
    }
    \addvspace{-4pt}
    \caption{Distribution of accuracy constraint slack across 42 benchmarks at $\delta = 0.05$. 
    Values falling below the dashed horizontal line represent violations of the accuracy target.}
    \label{fig:slack-distribution-vs-gamma}
\end{figure}

\myparagraph{Unsoundness of Prior Tuning Algorithms.}
The audit reveals that LOTUS and ScaleDoc systematically fail to meet their guarantees:
\begin{itemize}[leftmargin=1.5em]
    \item At nominal error budget $\delta = 0.05$, LOTUS is certified unsound on $80.4\%$ of all scenarios. At the strict $\gamma = 0.95$ target, LOTUS fails on $90.5\%$ of benchmarks \ifarxiv(\cref{tab:soundness_disaggregated_joint})\else \citep[Table~5]{actis_full}\fi.
    \item ScaleDoc is certified unsound on 39.9\% of scenarios at $\delta = 0.05$, and on 64.3\% of scenarios under $\gamma = 0.95$, with empirical failure rates reaching 100\% on multiple benchmarks.
    \item Tighter error budgets exacerbate these failures: at $\delta = 0.01$, LOTUS and ScaleDoc are certified unsound on 81.5\% and 53.0\% of scenarios, respectively.
\end{itemize}

\myparagraph{Severity of Target Violations.}
\cref{fig:slack-distribution-vs-gamma} plots the distribution of constraint slack across all 42 benchmarks with $\delta = 0.05$. 
For methods evaluated on joint precision and recall targets (ACTIS, BARGAIN, and LOTUS), slack is defined as $\min\{\Precision - \gamma, \, \Recall - \gamma\}$; for ScaleDoc, which is audited against its native F1 target, slack is defined as $F_1 - \gamma$.
Points below zero represent violations where deployed cascades fail to achieve the specified accuracy target. 
At an accuracy target of $\gamma = 0.95$, LOTUS's median delivered accuracy (the minimum of precision and recall) is 0.60, falling to 0.19 in the worst case. 
ScaleDoc's median delivered F1~score just falls short of the target at 0.945; however, it drops catastrophically to 0.09 in the worst case. 
These failures demonstrate that prior methods achieve low oracle invocation costs by choosing overly aggressive thresholds (see \ifarxiv\cref{app:soundness-analysis}\else App.~C of \citep{actis_full}\fi).

\myparagraph{Validity of ACTIS and BARGAIN.}
By contrast, ACTIS---under both importance sampling and uniform sampling, with asymptotic or non-asymptotic testing---maintains statistical validity. 
Across all 42 benchmarks and all tested configurations of $\gamma$ and $\delta$, ACTIS is never certified unsound. 
Its empirical maximum failure rate $\tilde{\delta}_{\max}$ never exceeds the nominal budget $\delta$. 
BARGAIN also achieves no certified unsound runs, establishing it as the only valid baseline against which cost efficiency can be fairly evaluated.

Crucially, the ablation of our safeguards (ACTIS w/o~safeguards) underscores why they are necessary when deploying asymptotic mixture tests in finite samples. 
Without safeguards, the empirical failure rate violates the nominal budget on up to $3.6\%$ of scenarios, with failure rates reaching up to $23.2\%$ at $\delta = 0.05$. 
These failures validate the theoretical vulnerabilities identified in \cref{sec:asymptotic-safeguards}: on highly selective queries where positives are rare, the asymptotic test is susceptible to premature termination on early success streaks and zero-variance sample draws. 
By incorporating safeguards, ACTIS maintains nominal error control even in the finite sample regime.

\begin{figure}
    \centering
    \addvspace{-4pt}
    \resizebox{0.85\linewidth}{!}{%
    \input{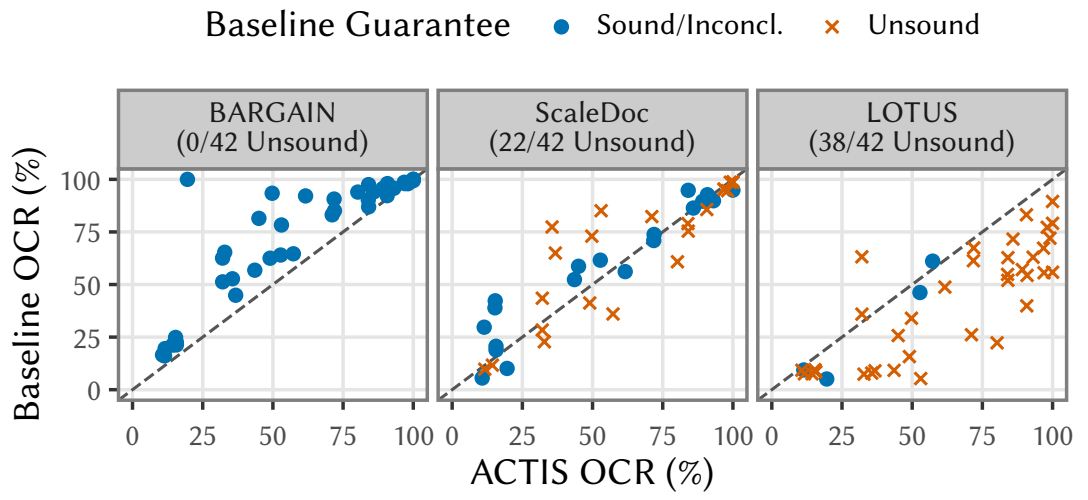}
    }
    \addvspace{-6pt}
    \caption{Pairwise comparison of oracle call rate (OCR) between baselines and ACTIS ($\gamma = 0.95, \delta = 0.05$). Crosses indicate workloads for which the baseline is certified unsound. ACTIS achieves a lower OCR on workloads for which the marker is above the dashed diagonal.}
    \label{fig:pairwise-head-to-head}
\end{figure}

\begin{table*}
\centering
\caption{Oracle call rate (\%) and tuner latency (ms) on representative benchmarks for precision\slash recall target $\gamma = 0.95$ and error budget $\delta = 0.05$. `Ideal' denotes an omniscient tuner that selects the minimum-cost valid thresholds using oracle labels for the workload. \badvalue{Red*} indicates certified unsoundness; boldface highlights the best result among statistically sound methods.}
\label{tab:representative_performance}
\label{tab:oracle_call_rates_representative}
\label{tab:tuning_latency_representative}
\footnotesize
\setlength{\tabcolsep}{4.2pt}
\addvspace{-4pt}
\begin{tabular}{llr rrrrrr rrrrr}
\toprule
\multicolumn{2}{c}{\textbf{Benchmark}} & & \multicolumn{6}{c}{\textbf{Oracle Call Rate (\%)}} & \multicolumn{5}{c}{\textbf{Tuner Latency (ms)}} \\
\cmidrule(lr){1-2} \cmidrule(lr){4-9} \cmidrule(lr){10-14}
\textbf{Workload} & \textbf{Proxy} & \textbf{$N$} & \textbf{Ideal} & \textbf{ACTIS} & \textbf{ACTIS-U} & \textbf{BARGAIN} & \textbf{ScaleDoc} & \textbf{LOTUS} & \textbf{ACTIS} & \textbf{ACTIS-U} & \textbf{BARGAIN} & \textbf{ScaleDoc} & \textbf{LOTUS} \\
\midrule
Onto & UFT & 11k & 6.2 & \textbf{32.1} & 61.3 & 62.6 & \badvalue{28.4\rlap{*}} & \badvalue{63.2\rlap{*}} & 17.2 & 42.6 & \textbf{11.2} & \badvalue{4.4\rlap{*}} & \badvalue{1.6\rlap{*}} \\
ImageNet & ResNet-50 & 50k & 0.0 & 19.6 & 100.0 & 100.0 & 10.1 & \textbf{5.1} & 48.1 & 191.6 & 123.2 & 7.1 & \textbf{6.2} \\
Jackson & ResNet-50 & 973k & 35.4 & \textbf{45.0} & 46.4 & 81.5 & 58.7 & \badvalue{25.7\rlap{*}} & 3854.3 & 10110.6 & 3580.0 & \textbf{109.6} & \badvalue{171.3\rlap{*}} \\
TACRED & SpanBERT & 23k & 6.5 & \textbf{32.1} & 54.4 & 51.3 & \badvalue{43.5\rlap{*}} & \badvalue{35.9\rlap{*}} & 43.7 & 69.1 & \textbf{22.2} & \badvalue{5.5\rlap{*}} & \badvalue{3.0\rlap{*}} \\
Court & GPT-5.6-luna & 1k & 44.5 & \textbf{80.2} & 88.1 & 94.0 & \badvalue{60.8\rlap{*}} & \badvalue{22.3\rlap{*}} & 3.6 & 5.6 & \textbf{2.2} & \badvalue{3.3\rlap{*}} & \badvalue{0.6\rlap{*}} \\
Review & GPT-5.6-luna & 10k & 10.2 & \textbf{32.8} & 51.7 & 65.3 & \badvalue{22.8\rlap{*}} & \badvalue{7.4\rlap{*}} & \textbf{10.7} & 18.4 & 12.3 & \badvalue{4.3\rlap{*}} & \badvalue{1.6\rlap{*}} \\
Screenplay & Llama-3.2-1B & 3k & 53.9 & \textbf{84.1} & 85.5 & 97.5 & 94.8 & \badvalue{52.0\rlap{*}} & 44.2 & 87.2 & 4.1 & \textbf{3.9} & \badvalue{0.9\rlap{*}} \\
Wiki & GPT-5.6-luna & 10k & 37.0 & 71.1 & \textbf{68.6} & 83.1 & \badvalue{82.3\rlap{*}} & \badvalue{26.1\rlap{*}} & 69.5 & 129.1 & \textbf{13.7} & \badvalue{3.4\rlap{*}} & \badvalue{1.6\rlap{*}} \\
PubMed Q0 & GPT-5.6-luna & 10k & 0.0 & \textbf{15.3} & 15.8 & 24.8 & 42.3 & \badvalue{9.4\rlap{*}} & \textbf{3.9} & 4.5 & 5.2 & 6.2 & \badvalue{1.8\rlap{*}} \\
BigPatent Q0\textsubscript{ext} & Llama-3.2-1B & 10k & 60.2 & 85.9 & \textbf{84.8} & 94.0 & 86.4 & \badvalue{71.6\rlap{*}} & 69.3 & 113.9 & 18.9 & \textbf{5.4} & \badvalue{1.9\rlap{*}} \\
GovReport Q10 & GPT-5.6-luna & 10k & 12.5 & \textbf{35.6} & 43.1 & 52.7 & \badvalue{77.4\rlap{*}} & \badvalue{7.9\rlap{*}} & 13.3 & 17.9 & \textbf{10.2} & \badvalue{4.6\rlap{*}} & \badvalue{1.9\rlap{*}} \\
\bottomrule
\end{tabular}
\end{table*}

\subsection{RQ2: End-to-End Cascade Efficiency} \label{sec:exp-efficiency}

Having established that LOTUS and ScaleDoc violate their guarantees, we evaluate the cost effectiveness of ACTIS against BARGAIN, the only statistically valid baseline. 
Recall that for this efficiency comparison, all methods are configured to guarantee the common precision and recall target $\gamma_P = \gamma_R = \gamma$ (setting ScaleDoc's F1 target to $\frac{2\gamma}{1+\gamma}$, as detailed in \cref{sec:exp-setup}). 
\cref{fig:pairwise-head-to-head} presents a pairwise head-to-head comparison of oracle call rates (OCR) across all 42~benchmarks for $\gamma = 0.95$ and $\delta = 0.05$. 
\cref{tab:oracle_call_rates_representative} provides a detailed snapshot across representative scenarios, with full results documented in \ifarxiv\cref{app:evaluation} (\cref{tab:all_benchmarks_performance_mean_only})\else \citep[Table~6]{actis_full}\fi.
In these tables, \emph{Ideal} denotes the minimum OCR that achieves the precision\slash recall target under perfect ground-truth knowledge. 
Valid methods (ACTIS and BARGAIN) incur a gap to this ideal because statistical certification inherently requires querying a minimum sample of oracle labels to guarantee the error budget $\delta$, even for accurate proxies (e.g., ImageNet and PubMed~Q0).

\myparagraph{Comparison Against BARGAIN.}
At $\gamma = 0.95$ and $\delta = 0.05$, ACTIS matches or reduces oracle call rate (OCR) versus BARGAIN on 39 of 42 benchmarks ($92.9\%$), and achieves the lowest OCR among all sound methods on 8 of the 11 benchmarks in \cref{tab:oracle_call_rates_representative}.
Even without importance sampling, our non-asymptotic uniform variant ACTIS-U matches or outperforms BARGAIN on 36 of 42 benchmarks \ifarxiv(\cref{tab:oracle_call_rates_all_benchmarks_mean_only})\else\citep[Table~6]{actis_full}\fi---e.g., reducing OCR from $81.5\%$ to $46.4\%$ on Jackson and from $65.3\%$ to $51.7\%$ on Review~Q0---demonstrating the gains from decoupled fixed-sequence testing and Value-of-Information stopping.
Furthermore, comparing ACTIS against ACTIS-U and BARGAIN on selective workloads highlights how importance sampling overcomes uniform sampling bottlenecks:
\begin{itemize}[leftmargin=1.5em]
    \item On ImageNet ($\gamma = 0.95$, $0.1\%$ prevalence), both BARGAIN and ACTIS-U exhaust the workload ($100\%$ OCR) starving for rare positives under uniform sampling, whereas ACTIS satisfies the guarantee at $19.6\%$ OCR ($5.1\times$ reduction).
    \item On Onto and TACRED (${\approx} 2.5\%$ prevalence), ACTIS cuts OCR nearly in half compared to both ACTIS-U ($32.1\%$ vs.\ $61.3\%$ and $54.4\%$) and BARGAIN ($62.6\%$ and $51.3\%$).
\end{itemize}

\myparagraph{Impact of Proxy Model Quality.}
Comparing cloud-based GPT-5.6-luna against local Llama-3.2-1B reveals clear architectural trade-offs. 
As shown in \ifarxiv\cref{tab:dataset-stats}\else Table~4 of~\citep{actis_full}\fi, GPT-5.6-luna produces higher PR-AUC and lower expected calibration error (ECE). 
Consequently, cascades using GPT-5.6-luna achieve a lower OCR (e.g., $80.2\%$ on Court~Q0 vs. $96.8\%$ for Llama-3.2-1B under $\gamma = 0.95$). 
ACTIS maintains validity under both proxies, letting operators trade off local compute against API spend.

\subsection{RQ3: Ablation and Sensitivity Analysis} \label{sec:exp-ablations}

We conduct ablation and sensitivity studies to analyze the core algorithmic components of ACTIS and characterize the impact of hyperparameters on cascade efficiency and tuning latency.

\begin{figure}
    \centering
    \resizebox{0.85\linewidth}{!}{%
    \input{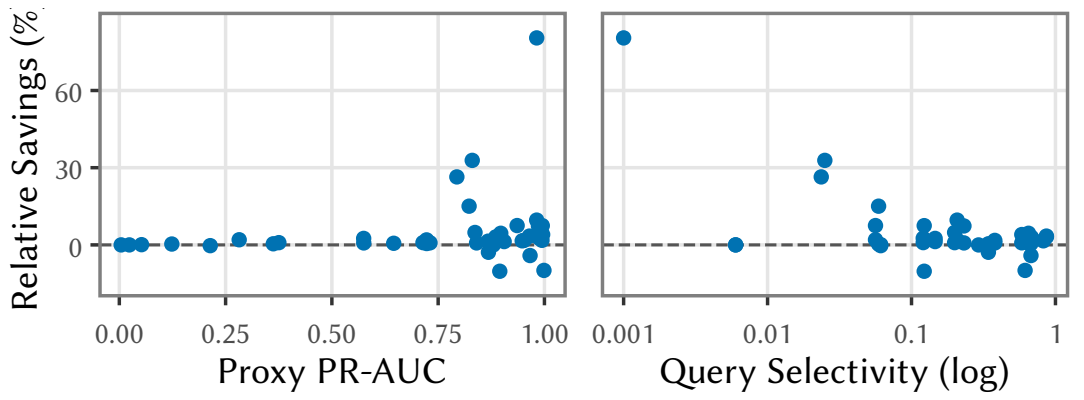}
    }
    \addvspace{-6pt}
    \caption{Relative reduction in oracle calls achieved by ACTIS with importance sampling versus uniform sampling across 42 benchmarks (with $\gamma = 0.95, \delta = 0.05$) as a function of proxy PR-AUC (left) and query positive rate $\Prevalence$ (right).}
    \label{fig:is_vs_uniform_pr_auc_selectivity}
\end{figure}

\myparagraph{Importance Sampling vs.\ Uniform Sampling.}
We evaluate the benefit of variance-optimal importance sampling by comparing default ACTIS against an identical configuration using uniform sampling (holding the safeguarded asymptotic testing engine fixed). 
At $\gamma = 0.95$ and $\delta = 0.05$, importance sampling achieves a lower oracle call rate on 39 of 42 benchmarks. 
As shown in \cref{fig:is_vs_uniform_pr_auc_selectivity}, efficiency gains are governed by two factors: proxy discriminative power and query selectivity. 
When proxy PR-AUC is high (left panel), $q^*$ accurately concentrates sampling mass on informative decision boundaries, tightening confidence sequences rapidly. 
Crucially, when target positives are rare ($\pi < 0.05$, right panel), uniform sampling wastes budget scanning uninformative true negatives, whereas $q^*$ oversamples high-scoring tuples to verify recall, reducing total oracle calls by up to 80\% relative to uniform sampling.

\begin{figure}
    \centering
    \resizebox{0.85\linewidth}{!}{%
    \input{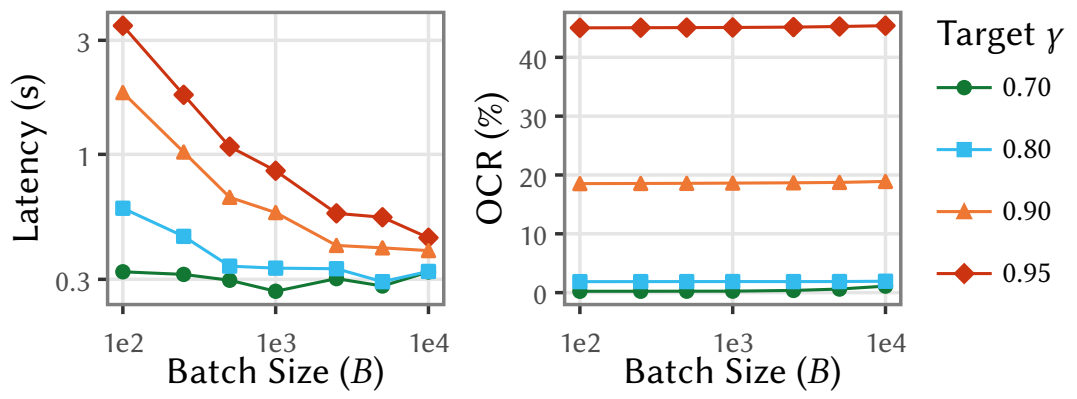}
    }
    \addvspace{-4pt}
    \caption{Sensitivity of ACTIS to batch size $B$ on Jackson ($N = 973{,}085, \delta = 0.05$) across accuracy targets $\gamma$. 
    At high accuracy, tuner latency (left) drops by up to $7\times$ as $B$ increases, while oracle call rate (right) remains virtually constant.}
    \label{fig:sensitivity-batch-size}
\end{figure}

\myparagraph{Batch Size Sensitivity.}
In sequential tuning, the batch size $B$ controls the granularity of oracle label acquisition and stopping rule evaluation. 
\cref{fig:sensitivity-batch-size} depicts the effect of scaling $B$ from 100 to $10{,}000$ on the million-row Jackson workload ($N = 973{,}085, \delta = 0.05$). 
As $B$ increases, tuning latency drops substantially (left panel)---falling from 3.45\,s to 0.55\,s at $B = 5{,}000$ ($6.3\times$ speedup) and 0.45\,s at $B = 10{,}000$ for $\gamma = 0.95$---because larger increments reduce the frequency of anytime-valid hypothesis tests. 
This speedup incurs negligible penalty in filtering efficiency (right panel): across all four accuracy targets, the oracle call rate increases by less than 0.38 percentage points (e.g., from 45.01\% to 45.38\% at $\gamma = 0.95$). 

This stability holds because $B$ remains small relative to the workload ($B / N \le 1\%$). 
By contrast, on smaller tables like PubMed Q0 ($N = 10{,}000$), coarse batch sizes such as $B = 5{,}000$ force the tuner to label half the dataset in a single step, artificially inflating the oracle call rate. 
Consequently, configuring $B$ as a small fraction of the table (e.g., $B \le 0.01 N$ or default $B = 100$) guarantees fine-grained stopping control while maintaining sub-second tuning latency across diverse workload scales.

\subsection{RQ4: Tuning Latency Overhead} \label{sec:exp-latency}

In a semantic database engine, cascade tuning acts as an online query optimization step that trades a small upfront computational investment during query planning for large downstream reductions in oracle execution. 
\cref{tab:tuning_latency_representative} reports the in-memory CPU time consumed by the tuning algorithms ($\gamma = 0.95, \delta = 0.05$), isolating algorithmic execution from external oracle I/O and network latency.

\myparagraph{Interactive Tuning Overhead.}
Across all 42 benchmarks at $\gamma = 0.95$, ACTIS requires a median CPU tuning time of only 27.3\,ms, completing in under 90\,ms on 41 of 42 benchmarks. 
This overhead is well within the budget of interactive query optimizers. 
While LOTUS and ScaleDoc exhibit faster runtimes (1.8\,ms and 5.2\,ms), their speed is an artifact of taking statistical shortcuts without multiple testing corrections or sequential validation (\cref{sec:exp-soundness}). 
Furthermore, the baseline runtimes reported in \cref{tab:representative_performance} reflect our optimized implementations; in their published upstream forms, both baselines exhibit substantial latency overheads. 
In a 10-trial benchmark on Review~Q0 ($N = 10\text{k}$), upstream LOTUS requires 560\,ms ($300 \times$ slower than our vectorized version at 1.9\,ms), while upstream BARGAIN requires 70\,ms ($5 \times$ slower than our vectorized version at 13.4\,ms). 
On the million-row Jackson benchmark, both upstream implementations fail to scale, timing out ($> 1$\,hour) on 80--100\% of trials. 
By contrast, ACTIS natively scales to million-row workloads in 3.45\,s (and 0.55\,s with $B = 5{,}000$), comparable to our vectorized BARGAIN baseline (3.58\,s).
Among statistically sound methods, ACTIS is the only framework that scales to million-row tables out-of-the-box within interactive latency bounds.

\myparagraph{Amortizing Tuning Costs.}
Directly measuring end-to-end latency across cloud LLM APIs is confounded by transient network jitter, throttling, and concurrency quotas. 
Since frontier LLMs like GPT-5.6-terra incur ${\approx} 0.9\text{ s}$ end-to-end latency per call (${\approx} 0.8\text{ s}$ time-to-first-token~\citep{aa_gpt56_terra}), oracle invocations dominate query runtime and cost, easily amortizing ACTIS's sub-second tuning overhead.
For example, on Review~Q0 ($N = 10{,}000$), ACTIS spends 10.7\,ms tuning, yet lowers the oracle call rate from 65.3\% (BARGAIN) to 32.8\%, avoiding 3,250 calls to GPT-5.6-terra. 
Even assuming high concurrency (e.g., 50 simultaneous streams), eliminating these 3,250 calls saves ${\approx} 59$\,s of query execution latency and reduces API expenditure (proxy plus oracle) by 36\% (\$1.60 vs. \$2.50), amortizing the 10.7\,ms tuning phase by four orders of magnitude. 
On the million-row Jackson benchmark, ACTIS requires 3.45\,s with $B = 100$ (or 0.55\,s with $B = 5{,}000$), but eliminates over 355,000 oracle calls compared to BARGAIN, potentially saving hours of downstream GPU compute.

\section{Related Work} \label{sec:related}

\myparagraph{Model Cascades.}
Model cascades~\citep{viola2001rapid} were introduced to database query processing to accelerate expensive machine learning predicates over unstructured data~\citep{kang2017noscope,hsieh2018focus,lu2018accelerating}. 
To address the lack of accuracy guarantees in early systems, \citet{kang2020approximate} proposed SUPG, which uses importance sampling and asymptotic confidence bounds to tune a single-threshold cascade for either a precision \emph{or} recall target. 
\citet{zeighami2025cut} showed that SUPG's asymptotic guarantees miss targets at small sample sizes and proposed BARGAIN, which provides finite-sample guarantees for single-threshold cascades using non-asymptotic betting martingales.

Recent systems use \emph{two-threshold} cascades to filter negatives and accept positives simultaneously. 
LOTUS~\citep{patel2025semantic} extends SUPG to joint precision and recall targets, while SUPG-IT~\citep{liskowski2026streaming} adapts LOTUS's tuning algorithm to streaming batches. 
ScaleDoc~\citep{zhang2026scaledoc} tunes a two-threshold cascade for a target F1~score using proportional stratified sampling, histogram smoothing, and Pareto-frontier threshold search. 
Despite claiming statistical guarantees, LOTUS, SUPG-IT, and ScaleDoc rely on analyses containing errors or heuristic approximations \ifarxiv(see \cref{app:soundness-analysis})\else\citep[App.~C]{actis_full}\fi; our audit finds high failure rates for LOTUS and ScaleDoc (\cref{sec:exp-soundness}). 
A recent, unpublished code addition by BARGAIN's authors, BARGAIN-PR~\citep{shankar2026bargainpr}, provides finite-sample validity for two-threshold cascades under a shared target ($\gamma_P = \gamma_R$), but relies on uniform and contiguous-window sampling, tests a conservative upper-tail precision condition, and lacks cost-aware stopping. 
By contrast, ACTIS supports distinct $(\gamma_P, \gamma_R)$ targets and combines anytime-valid sequential testing, decoupled fixed-sequence search, variance-minimizing importance sampling, and Value-of-Information stopping.

\myparagraph{Proxy Models for Cascades.}
Cascade efficiency depends on the discriminative power of its proxy scorer. 
Prior work has explored diverse proxy architectures, including lightweight classifiers over bag-of-words or domain features~\citep{lu2018accelerating,yang2022optimizing}, task-agnostic embedding indexes~\citep{kang2022tasti}, small LLM token log-probabilities~\citep{patel2025semantic}, and classifiers trained on precomputed vector embeddings~\citep{chung2026cost}. 
ScaleDoc~\citep{zhang2026scaledoc} trains a query-aware multilayer perceptron over precomputed LLM document embeddings using two-phase contrastive learning to encourage smooth, monotonic, and bipolar score distributions. 
These techniques are orthogonal to ACTIS, which treats the proxy scorer as a black box and adapts to its calibration quality.

\myparagraph{Beyond Two-Threshold Cascades.}
Recent work explores richer optimization spaces and non-cascade paradigms for semantic filtering. 
\citet{shankar2026task} introduce \emph{task cascades}, using an LLM agent to assemble cascades that vary the model, operation, and document portion processed at each stage. 
Stretto~\citep{sanmartino2026stretto} jointly optimizes model size, KV-cache compression, and cascade thresholds via gradient descent over a continuous plan relaxation, using marginal Bayesian credible intervals to bound precision and recall. 
Beyond cascades, \citet{hou2026linear} propose Clustering-Sampling-Voting (CSV), which clusters tuple embeddings, queries the oracle on cluster samples, and propagates labels through voting with recursive re-clustering. 
However, CSV bounds only per-cluster accuracy, which is uninformative under class imbalance, while Stretto and CSV omit statistical corrections for adaptive search and partitioning \ifarxiv(see \cref{app:soundness-analysis})\else\citep[App.~C]{actis_full}\fi.

\myparagraph{Semantic Operators and Data Processing Systems.}
While ACTIS optimizes semantic filters in isolation, parallel work provides specialized approximations with guarantees for other semantic operators, including stratified sampling for approximate aggregations~\citep{kang2019blazeit,kang2021accelerating} and featurized decomposition for semantic joins~\citep{zeighami2026featurized}. 
These operators also serve as primitives in declarative semantic data systems~\citep{patel2025semantic,liu2025palimpzest,shankar2025docetl,dai2024uqe,lin2025querying,he2025sql,liskowski2026cortex}, which use holistic query optimizers for cost-based model and operator selection~\citep{liu2025palimpzest,russo2026abacus,huang2026scope} or agentic query rewriting~\citep{shankar2025docetl,wei2026multi}. 
ACTIS complements these system-level optimizers by providing a drop-in physical filter operator with statistical accuracy guarantees.

\myparagraph{Routing and Cascades in Machine Learning.}
Outside databases, query routers~\citep{ong2025routellm,ding2024hybrid} and sequential multi-model cascades~\citep{chen2024frugalgpt} route inputs across LLMs of varying cost and capability, typically optimizing empirical cost-quality trade-offs without statistical guarantees. 
In machine learning, Learn-then-Test~\citep{angelopoulos2025learn} calibrates post-processing thresholds under finite-sample risk control using fixed-sample $p$-values and FWER-controlling fixed-sequence testing over a pre-labeled i.i.d.\ calibration set. 
Unlike offline calibration, semantic query processing acquires labels online from an expensive oracle, which ACTIS minimizes via anytime-valid testing, importance sampling, and cost-aware stopping.

\section{Conclusion} \label{sec:conclusion}

In this paper, we audited the statistical soundness of existing model cascade tuning methods for semantic filters, revealing that state-of-the-art techniques frequently violate their claimed precision and recall guarantees. 
To deliver dependable guarantees efficiently, we introduced ACTIS, which combines anytime-valid sequential testing, runtime safeguards, decoupled fixed-sequence search, variance-minimizing importance sampling, and Value-of-Information early stopping. 
Across 42 benchmarks spanning cloud and local proxies, ACTIS never violates the user-specified failure budget, matches or outperforms the only statistically valid baseline on 39 of 42 settings, and reduces oracle invocations by up to $5.1\times$.
More broadly, our findings underscore a critical lesson for semantic data processing systems: to earn the trust expected of database engines, probabilistic accuracy guarantees must withstand rigorous empirical auditing rather than fail silently in practice.

\begin{acks}
 This research was supported by the Australian Government through the Office of National Intelligence's funded National Intelligence Discovery Grant program (ID250100316). The views expressed herein are those of the authors and are not necessarily those of the Australian Government or the Office of National Intelligence.
 This research was also supported by The University of Melbourne's Research Computing Services and the Petascale Campus.
\end{acks}

\printbibliography
\ifarxiv
\FloatBarrier
\appendix

\section{Technical Details for Section~\ref{sec:method}} \label[appendix]{app:proofs}

This appendix provides the mathematical derivations, hyperparameter configurations, and formal proofs supporting the ACTIS framework in \cref{sec:method}. 
Specifically, \cref{app:rescaling} details the affine transformations used to map importance-weighted recall and precision increments onto $[0, 1]$ for the non-asymptotic betting test (\cref{sec:single-threshold-test}). 
\cref{app:test-hyperparameters} specifies the hyperparameter settings for both anytime-valid testing subroutines, including the proxy-surrogate horizon variance estimator and null-boundary error clamping for the asymptotic Gaussian mixture test. 
\cref{app:safeguards} derives the single-step variance floor (Safeguard~1) and the instrumental-adjusted null failure probability (Safeguard~3) introduced in \cref{sec:asymptotic-safeguards}. 
\Cref{app:monotonicity-proof,app:fwer-proof} provide the formal proofs of \cref{prop:monotonicity,prop:fwer-control}, respectively. 
\Cref{app:stopping-details} specifies the warmup threshold $n_{\min}$ used by the Value-of-Information stopping rule (\cref{sec:stopping-condition}). 
Finally, \cref{app:opt-instrumental} formalizes the joint variance objective, proves \cref{prop:opt-instrumental}, and derives the variance-minimizing component weights in \cref{sec:instrumental-design}.

\subsection{Rescaling for the Non-Asymptotic Betting-Based Test} \label[appendix]{app:rescaling}

When implementing $\Call{RecallMet}{}$ and $\Call{PrecisionMet}{}$ using non-asymptotic betting martingales (\cref{sec:non-asymptotic-test}), test increments must lie within the unit interval $[0, 1]$.
However, the raw recall and precision increments $f_R$ and $f_P$ (\cref{eqn:recall-increment,eqn:precision-increment}) violate this condition: they take negative values on constraint violations (false negatives and false positives, respectively) and can exceed 1 under non-uniform importance weights.
We therefore map each raw increment $Z \in \{f_R, f_P\}$ into $[0, 1]$ via an affine transformation $Z \mapsto (Z + \text{shift}) / \text{scale}$, which shifts the corresponding null hypothesis threshold from 0 to $m \coloneqq \text{shift} / \text{scale} \in (0, 1)$.

To define shift and scale parameters that guarantee $[0, 1]$ support across the workload $\D$, we establish upper bounds on the importance weights:
\begin{equation}
    \wmax^{\ge}(\tau) \coloneqq \max_{x \in \D \colon \proxy(x) \ge \tau} w(x), \quad
    \wmax^{<}(\tau) \coloneqq \max_{x \in \D \colon \proxy(x) < \tau} w(x).
    \label{eqn:importance-weight-bounds}
\end{equation}
Under uniform sampling, $w(x) \equiv 1$ yielding $\wmax^{\ge}(\tau) = \wmax^{<}(\tau) = 1$ for all thresholds.

For recall, the increment $f_R(X, Y; \lo{\tau})$ is bounded from below by $-\gamma_R \wmax^{<}(\lo{\tau})$ and bounded from above by $(1 - \gamma_R) \wmax^{\ge}(\lo{\tau})$. 
Defining 
\begin{align*}
    \text{shift}_R(\lo{\tau}) &\coloneqq \gamma_R \wmax^{<}(\lo{\tau}) \\
    \text{scale}_R(\lo{\tau}) &\coloneqq (1 - \gamma_R) \wmax^{\ge}(\lo{\tau}) + \gamma_R \wmax^{<}(\lo{\tau})
\end{align*}
ensures $[0, 1]$-bounded increments, which are evaluated against the rescaled target margin $m_R(\lo{\tau}) \coloneq \text{shift}_R(\lo{\tau}) / \text{scale}_R(\lo{\tau})$.

Similarly, for precision, the increment $f_P(X, Y; \lo{\tau}, \up{\tau})$ is bounded from below by $-\gamma_P \wmax^{\ge}(\up{\tau})$ and bounded from above by $(1 - \gamma_P) \wmax^{\ge}(\lo{\tau})$. 
Defining 
\begin{align*}
    \text{shift}_P(\up{\tau}) &\coloneqq \gamma_P \wmax^{\ge}(\up{\tau}) \\
    \text{scale}_P(\lo{\tau}, \up{\tau}) &\coloneqq (1 - \gamma_P) \wmax^{\ge}(\lo{\tau}) + \gamma_P \wmax^{\ge}(\up{\tau})
\end{align*}
ensures $[0, 1]$-bounded increments, which are evaluated against the rescaled target margin $m_P(\lo{\tau}, \up{\tau}) \coloneqq \text{shift}_P(\up{\tau}) / \text{scale}_P(\lo{\tau}, \up{\tau})$.

\subsection{Hyperparameter Settings for Anytime-Valid Tests} \label[appendix]{app:test-hyperparameters}

This section details how hyperparameters are configured for the two anytime-valid test subroutines in Section~\ref{sec:single-threshold-test}.

\subsubsection{Non-Asymptotic Betting-Based Test} \label[appendix]{app:hyperparams-betting}

For the betting test, predictable betting fractions $\lambda_i$ in \cref{eqn:betting-fraction} are evaluated using the standard default hyperparameter settings recommended by \citet{waudby2024estimating}:
\begin{itemize}[leftmargin=2em]
    \item Truncation scale parameter $c = 1/2$.
    \item Prior mean $\tilde{\mu}_0 = 1/2$ and prior variance $\tilde{\sigma}_0^2 = 1/4$ for the regularized sample variance estimator $\tilde{\sigma}_t^2$, initialized with one pseudo-observation.
\end{itemize}

\subsubsection{Asymptotic Gaussian Mixture Test} \label[appendix]{app:hyperparams-mixture}

To optimize statistical power over a planned tuning horizon $n_0$, the baseline prior variance parameter $v$ in \cref{eqn:gauss-mix-supermartingale} is set to the estimated cumulative process variance $v^{\text{est}} \coloneqq n_0 \tilde{\sigma}^2$ with $n_0 = \min(1000, \max(50, N / 10))$.
The critical operating point for Type~I error control is the boundary of the null hypothesis where $\Delta_R(\lo{\tau}) = 0$ or $\Delta_P(\lo{\tau}, \up{\tau}) = 0$. 
In the interior of the null ($\Delta < 0$), drift is negative, impeding rejection; in the alternative ($\Delta > 0$), positive drift accelerates rejection. 
We therefore evaluate the process variance $\sigma^2$ specifically at the null boundary.

\begin{proposition}[Process Variance at the Boundary Null] \label[proposition]{prop:var-null-boundary}
At the boundary of the null hypotheses $H_{0,R}(\lo{\tau})$ and $H_{0,P}(\lo{\tau}, \up{\tau})$, the process variances are given by:
\begin{align}
    \sigma_R^2(\lo{\tau}) &= (1 - \gamma_R)^2 \TPR_w(\lo{\tau}) + \gamma_R^2 \FNR_w(\lo{\tau}), \label{eqn:app-sigma-R} \\
    \sigma_P^2(\lo{\tau}, \up{\tau}) &= (1 - \gamma_P)^2 \TPR_w(\lo{\tau}) + \gamma_P^2 \FPR_w(\up{\tau}), \label{eqn:app-sigma-P}
\end{align}
where the importance-weighted true positive, false negative, and false positive components are defined as:
\begin{align*}
    \TPR_w(\lo{\tau}) &\coloneqq \E_{X \sim q, Y}[ w(X)^2 Y \indicator{\proxy(X) \ge \lo{\tau}} ], \\
    \FNR_w(\lo{\tau}) &\coloneqq \E_{X \sim q, Y}[ w(X)^2 Y \indicator{\proxy(X) < \lo{\tau}} ], \\
    \FPR_w(\up{\tau}) &\coloneqq \E_{X \sim q, Y}[ w(X)^2 (1 - Y) \indicator{\proxy(X) \ge \up{\tau}} ].
\end{align*}
\end{proposition}

\begin{proof}
For brevity, we let $f = f_R(X, Y; \lo{\tau})$ and $f = f_P(X, Y; \lo{\tau}, \up{\tau})$ denote the increments for $H_{0, R}(\lo{\tau})$ and $H_{0, P}(\lo{\tau}, \up{\tau})$ respectively.
At the boundary of the null hypothesis, the expected increment is zero ($\E_q[f] = 0$). 
The variance reduces to the uncentered second moment: $\var_q(f) = \E_q[f^2]$. 
Since oracle labels and indicator functions are $\{0, 1\}$-valued, we have $Y^2 = Y$, $(1 - Y)^2 = 1 - Y$, and $\indicator{\cdot}^2 = \indicator{\cdot}$.

For recall, squaring the increment \cref{eqn:recall-increment} yields:
\begin{align*}
    f_R(X, Y; \lo{\tau})^2 &= w(X)^2 Y^2 \left( \indicator{\proxy(X) \ge \lo{\tau}} - \gamma_R \right)^2 \\
    &= w(X)^2 Y \left( (1 - \gamma_R)^2 \indicator{\proxy(X) \ge \lo{\tau}} + \gamma_R^2 \indicator{\proxy(X) < \lo{\tau}} \right).
\end{align*}
Taking expectations with respect to $X \sim q$ and $Y \sim \oracle(X)$ yields \cref{eqn:app-sigma-R}.

For precision, note that because $\up{\tau} \ge \lo{\tau}$, the event $\proxy(X) \ge \up{\tau}$ implies $\proxy(X) \ge \lo{\tau}$. 
Moreover, an item cannot be simultaneously positive and negative ($Y(1 - Y) = 0$), so cross-product terms vanish. 
Squaring the precision increment \cref{eqn:precision-increment} yields:
\begin{align*}
    f_P(X, Y; \lo{\tau}, \up{\tau})^2 &= w(X)^2 \big( (1 - \gamma_P)^2 Y \indicator{\proxy(X) \ge \lo{\tau}} \\
    & \qquad {} + \gamma_P^2 (1 - Y) \indicator{\proxy(X) \ge \up{\tau}} \big).
\end{align*}
Taking expectations under $X \sim q$ and $Y \sim \oracle(X)$ yields \cref{eqn:app-sigma-P}.
\end{proof}

\myparagraph{Proxy Surrogate Variance with Null Boundary Clamping.}
Prior to cascade tuning, ground-truth oracle labels are unobserved. 
We estimate the variance components across the workload $\D$ by substituting proxy confidence scores $\proxy(x)$ as surrogates for oracle expectations $\E[Y \mid x]$:
\begin{align}
    \widetilde{\TPR}_w(\lo{\tau}) &\coloneqq \frac{1}{N} \sum_{x \in \D} w(x) \proxy(x) \indicator{\proxy(x) \ge \lo{\tau}}, \\
    \widetilde{\FNR}_w(\lo{\tau}) &\coloneqq \frac{1}{N} \sum_{x \in \D} w(x) \proxy(x) \indicator{\proxy(x) < \lo{\tau}}, \\
    \widetilde{\FPR}_w(\up{\tau}) &\coloneqq \frac{1}{N} \sum_{x \in \D} w(x) (1 - \proxy(x)) \indicator{\proxy(x) \ge \up{\tau}}.
\end{align}
If the proxy model is miscalibrated, raw surrogate error estimates $\widetilde{\FNR}_w$ or $\widetilde{\FPR}_w$ could artificially inflate or deflate the estimated variance. 
To improve robustness, we constrain these error terms to be consistent with the theoretical null boundary:

\begin{proposition}[Null Boundary Error Bounds] \label[proposition]{prop:variance_upper_bounds}
At the boundary of the null hypotheses $H_{0, R}(\lo{\tau})$ and $H_{0, P}(\lo{\tau}, \up{\tau})$, the importance-weighted error components satisfy:
\begin{align}
    \FNR_w(\lo{\tau}) &\le \rho_R(\lo{\tau}) \left(\frac{1 - \gamma_R}{\gamma_R}\right) \TPR_w(\lo{\tau}), \label{eqn:app-fnr-bound} \\
    \FPR_w(\up{\tau}) &\le \rho_P(\lo{\tau}, \up{\tau}) \left(\frac{1 - \gamma_P}{\gamma_P}\right) \TPR_w(\lo{\tau}), \label{eqn:app-fpr-bound}
\end{align}
where the partition weight disparity ratios are:
\begin{equation*}
    \rho_R(\lo{\tau}) \coloneqq \frac{\max_{x \colon \proxy(x) < \lo{\tau}} w(x)}{\min_{x \colon \proxy(x) \ge \lo{\tau}} w(x)}, \quad
    \rho_P(\lo{\tau}, \up{\tau}) \coloneqq \frac{\max_{x \colon \proxy(x) \ge \up{\tau}} w(x)}{\min_{x \colon \proxy(x) \ge \lo{\tau}} w(x)}.
\end{equation*}
\end{proposition}

\begin{proof}
At the null boundary for recall, $\Delta_R(\lo{\tau}) = \TPR(\lo{\tau}) - \gamma_R \Prevalence = 0$. 
Since $\Prevalence = \TPR(\lo{\tau}) + \FNR(\lo{\tau})$, this implies $\FNR(\lo{\tau}) = \frac{1 - \gamma_R}{\gamma_R} \TPR(\lo{\tau})$. 
Bounding the weighted components:
\begin{align*}
    \FNR_w(\lo{\tau}) &= \E_{X \sim p, Y}[w(X) Y \indicator{\proxy(X) < \lo{\tau}}] \le \Bigl(\max_{x \colon \proxy(x) < \lo{\tau}} w(x)\Bigr) \FNR(\lo{\tau}), \\
    \TPR_w(\lo{\tau}) &= \E_{X \sim p, Y}[w(X) Y \indicator{\proxy(X) \ge \lo{\tau}}] \ge \Bigl(\min_{x \colon \proxy(x) \ge \lo{\tau}} w(x)\Bigr) \TPR(\lo{\tau}).
\end{align*}
Combining these inequalities with the null boundary relation yields \cref{eqn:app-fnr-bound}. 
An identical derivation holds for precision, establishing \cref{eqn:app-fpr-bound}.
\end{proof}

Clamping the surrogate error components by their theoretical null limits yields the process variance estimators:
\begin{align*}
    \widetilde{\sigma}_R^2(\lo{\tau}) &\coloneqq (1 - \gamma_R)^2 \widetilde{\TPR}_w(\lo{\tau}) \\
    &\quad {} + \gamma_R^2 \min\left( \widetilde{\FNR}_w(\lo{\tau}), \; \rho_R(\lo{\tau}) \left(\frac{1 - \gamma_R}{\gamma_R}\right) \widetilde{\TPR}_w(\lo{\tau}) \right), \\
    \widetilde{\sigma}_P^2(\lo{\tau}, \up{\tau}) &\coloneqq (1 - \gamma_P)^2 \widetilde{\TPR}_w(\lo{\tau}) \\
    &\quad {} + \gamma_P^2 \min\left( \widetilde{\FPR}_w(\up{\tau}), \; \rho_P(\lo{\tau}, \up{\tau}) \left(\frac{1 - \gamma_P}{\gamma_P}\right) \widetilde{\TPR}_w(\lo{\tau}) \right).
\end{align*}
The baseline estimated horizon variances are then $v_R^{\text{est}} = n_0 \widetilde{\sigma}_R^2(\lo{\tau})$ and $v_P^{\text{est}} = n_0 \widetilde{\sigma}_P^2(\lo{\tau}, \up{\tau})$. 
When Safeguard~1 is active (\cref{sec:asymptotic-safeguards}), these are floored via \cref{eqn:v-final} during cascade tuning.

\subsection{Finite-Sample Safeguard Derivations} \label[appendix]{app:safeguards}

This section provides derivations for the runtime safeguards introduced in Section~\ref{sec:asymptotic-safeguards}.

\subsubsection{Proof of Single-Step Boundary Safety (Safeguard 1)}

\begin{proposition}[Single-Step Boundary Safety] \label[proposition]{prop:single_step_safety}
    Let $M_t$ be the Gaussian mixture process in \cref{eqn:gauss-mix-supermartingale} with rejection threshold $1/\alpha$ for significance level $\alpha \in (0, 1)$. 
    Suppose the increments $Z_t$ are bounded from above by $b$ and the prior variance satisfies:
    \begin{equation*}
        v \ge v^{\text{floor}} \coloneqq \frac{2 b^2}{\log(1 / \alpha)}.
    \end{equation*}
    For any first observation $Z_1 \le b$ (at sample size $t = 1$, where $V_1 = 0$), the accumulated wealth satisfies:
    \begin{equation*}
        M_1 \le \left(\frac{1}{\alpha}\right)^{1/4} < \frac{1}{\alpha}.
    \end{equation*}
    Consequently, no single observation can ever cross the rejection boundary on the first sample.
\end{proposition}

\begin{proof}
    At sample size $t = 1$, the cumulative sample mean is $\bar{Z}_1 = Z_1$. 
    The empirical variance accumulator evaluates to:
    \begin{equation*}
        V_1 = (Z_1 - \bar{Z}_1)^2 = 0,
    \end{equation*}
    and the cumulative drift evaluates to $S_1 = Z_1$ (assuming $m = 0$). 
    If $Z_1 \le 0$, the process yields $M_1 = 0$. 
    For any positive increment $Z_1 \in (0, b]$, substituting $V_1 = 0$ into \cref{eqn:gauss-mix-supermartingale} yields
    $M_1 = \exp( Z_1^2/ (2 v) )$.
    Since $Z_1 \le b$ and $v \ge 2 b^2 / \log(1/\alpha)$, we obtain:
    \begin{equation*}
        M_1 \le \exp\left( \frac{b^2}{2 \cdot \frac{2 b^2}{\log(1/\alpha)}} \right) = \exp\left( \frac{\log(1/\alpha)}{4} \right) = \left(\frac{1}{\alpha}\right)^{1/4} < \frac{1}{\alpha}.
    \end{equation*}
\end{proof}

\subsubsection{Derivation of Instrumental-Adjusted Failure Probability (Safeguard 3)}

\begin{proposition}[Instrumental-Adjusted Null Failure Probability] \label[proposition]{prop:instrumental-adjusted-p0}
Let $F, S \subset \D \times \{0, 1\}$ denote disjoint failure and success subsets, and let $A \coloneqq F \cup S$ denote the active subset.
Let $\pi_F \coloneqq \E_p[\indicator{(X, Y) \in F}]$ and $\pi_S \coloneqq \E_p[\indicator{(X, Y) \in S}]$ denote their population masses under nominal uniform distribution $p$.
Define the average instrumental densities over the failure and success regions:
\begin{align*}
    \bar{q}_F &\coloneqq \frac{\sum_{x \in \D} q(x) \E[\indicator{(x, Y) \in F} \mid x]}{N \pi_F}, \\
    \bar{q}_S &\coloneqq \frac{\sum_{x \in \D} q(x) \E[\indicator{(x, Y) \in S} \mid x]}{N \pi_S}.
\end{align*}
Under the boundary null hypothesis $\pi_F / \pi_S = (1 - \gamma) / \gamma$, the conditional probability of drawing a failure under instrumental distribution $q$ given that an active record is sampled is:
\begin{equation}
    p_0 \coloneqq \Pr_{X \sim q, Y}\bigl[(X, Y) \in F \bigm| (X, Y) \in A\bigr] = \frac{(1 - \gamma) \eta}{\gamma + (1 - \gamma) \eta},
    \label{eqn:app-adjusted-p0}
\end{equation}
where $\eta \coloneqq \bar{q}_F / \bar{q}_S$ is the instrumental density ratio.
\end{proposition}

\begin{proof}
Applying Bayes' rule and the definition of conditional probability:
\begin{align*}
    p_0 &= \frac{\Pr_q[(X, Y) \in F]}{\Pr_q[(X, Y) \in F] + \Pr_q[(X, Y) \in S]} \\
        &= \frac{\bar{q}_F \pi_F}{\bar{q}_F \pi_F + \bar{q}_S \pi_S} 
        = \frac{(\bar{q}_F / \bar{q}_S) (\pi_F / \pi_S)}{1 + (\bar{q}_F / \bar{q}_S) (\pi_F / \pi_S)}.
\end{align*}
Substituting the boundary null condition $\pi_F / \pi_S = (1 - \gamma) / \gamma$ and instrumental density ratio $\eta = \bar{q}_F / \bar{q}_S$ yields \cref{eqn:app-adjusted-p0}.
\end{proof}

\myparagraph{Instrumental Ratio for Recall ($\eta_R$).}
For recall at threshold $\lo{\tau}$, failures (false negatives) reside strictly in the proxy rejection partition $\D_0(\lo{\tau}) = \{x \colon \proxy(x) < \lo{\tau}\}$, while successes (true positives) reside in the cascade acceptance partition $\D \setminus \D_0(\lo{\tau}) = \{x \colon \proxy(x) \ge \lo{\tau}\}$. 
Prior to cascade tuning, the conditional label prevalence $\E[Y \mid x]$ within each partition is unobserved.
Under an uninformative prior that treats conditional label density as homogeneous within each partition, the unknown constants factor out of the numerator and denominator sums.
This reduces $\bar{q}_F$ and $\bar{q}_S$ to the unweighted arithmetic means of $q(x)$ across the respective partitions:
\begin{equation*}
    \eta_R(\lo{\tau}) \coloneqq \frac{|\D_0(\lo{\tau})|^{-1} \sum_{x \in \D_0(\lo{\tau})} q(x)}{|\D \setminus \D_0(\lo{\tau})|^{-1} \sum_{x \in \D \setminus \D_0(\lo{\tau})} q(x)}.
\end{equation*}
Substituting $\eta_R(\lo{\tau})$ into \cref{eqn:app-adjusted-p0} with $\gamma = \gamma_R$ yields the adjusted null failure probability $p_{0, R}(\lo{\tau})$.

\myparagraph{Instrumental Ratio for Precision ($\eta_P$).}
For precision at $(\lo{\tau}, \up{\tau})$, failures (false positives) and successes (true positives) can both occur throughout the cascade acceptance region. 
We use the proxy scores as a surrogate for the oracle probability: an item with proxy score $\proxy(x)$ has surrogate false-positive probability $1 - \proxy(x)$ and surrogate true-positive probability $\proxy(x)$.
Weighting instrumental draws by these proxy surrogates gives:
\begin{align*}
    \bar{q}_{\text{FP}}(\up{\tau}) &\coloneqq \frac{\sum_{x \colon \proxy(x) \ge \up{\tau}} q(x) (1 - \proxy(x))}{\sum_{x \colon \proxy(x) \ge \up{\tau}} (1 - \proxy(x))}, \\
    \bar{q}_{\text{TP}}(\lo{\tau}) &\coloneqq \frac{\sum_{x \colon \proxy(x) \ge \lo{\tau}} q(x) \proxy(x)}{\sum_{x \colon \proxy(x) \ge \lo{\tau}} \proxy(x)}.
\end{align*}
To guard against proxy miscalibration, we clamp the ratio by the extremal instrumental density range over the active regions:
\begin{align*}
    \eta_{P, \min}(\lo{\tau}, \up{\tau}) &\coloneqq \frac{\min_{x \colon \proxy(x) \ge \up{\tau}} q(x)}{\max_{x \colon \proxy(x) \ge \lo{\tau}} q(x)}, \\
    \eta_{P, \max}(\lo{\tau}, \up{\tau}) &\coloneqq \frac{\max_{x \colon \proxy(x) \ge \up{\tau}} q(x)}{\min_{x \colon \proxy(x) \ge \lo{\tau}} q(x)}, \\
    \eta_P(\lo{\tau}, \up{\tau}) &\coloneqq \operatorname{clip}\left( \frac{\bar{q}_{\text{FP}}(\up{\tau})}{\max(\bar{q}_{\text{TP}}(\lo{\tau}), 10^{-12})}, \; \eta_{P, \min}, \; \eta_{P, \max} \right).
\end{align*}
Evaluating \cref{eqn:app-adjusted-p0} with $\gamma = \gamma_P$ and clamping the result to $p_{0, P} \coloneqq \min\bigl(0.50, \, \frac{(1 - \gamma_P)\eta_P}{\gamma_P + (1 - \gamma_P)\eta_P}\bigr)$ yields the precision null failure probability. 
The 0.5 ceiling prevents a miscalibrated proxy from asserting that cascade positive predictions are more likely to fail than succeed under the null ($\eta_P \ge \gamma_P / (1 - \gamma_P)$, e.g., $\ge 9$ for $\gamma_P \geq 0.9$).

\subsection{Proof of Proposition~\ref{prop:monotonicity}} \label[appendix]{app:monotonicity-proof}

    Consider the true positive rate $\TPR(\lo{\tau}) = \E[Y \indicator{\proxy(X) \ge \lo{\tau}}]$.
    For any $\lo{\tau} \le \lo{\tau}'$, the indicator satisfies $\indicator{\proxy(X) \ge \lo{\tau}} \ge \indicator{\proxy(X) \ge \lo{\tau}'}$ almost surely. 
    Since $Y \ge 0$, we have $Y \indicator{\proxy(X) \ge \lo{\tau}} \ge Y \indicator{\proxy(X) \ge \lo{\tau}'}$, which implies $\TPR(\lo{\tau}) \ge \TPR(\lo{\tau}')$. 
    Subtracting the constant $\gamma_R \Prevalence$ preserves the inequality, yielding $\Delta_R(\lo{\tau}) \ge \Delta_R(\lo{\tau}')$, proving~(i). 

    Now consider the false positive rate $\FPR(\up{\tau}) = \E[(1 - Y) \indicator{\proxy(X) \ge \up{\tau}}]$.
    Following an identical argument, $1 - Y \ge 0$ implies $\FPR(\up{\tau}) \ge \FPR(\up{\tau}')$ for any $\up{\tau} \le \up{\tau}'$. 
    Multiplying by $-\gamma_P < 0$ reverses the inequality: $-\gamma_P \FPR(\up{\tau}) \le -\gamma_P \FPR(\up{\tau}')$. 
    Adding the constant $(1 - \gamma_P) \TPR(\lo{\tau})$ yields $\Delta_P(\lo{\tau}, \up{\tau}) \le \Delta_P(\lo{\tau}, \up{\tau}')$, completing the proof of~(ii).

\subsection{Proof of Proposition~\ref{prop:fwer-control}} \label[appendix]{app:fwer-proof}

    A joint constraint violation occurs if $\lo{\tau}^\star$ violates recall or $\up{\tau}^\star$ violates precision:
    \begin{equation*}
        \text{FWER} \le \Pr[\lo{\tau}^\star \text{ violates recall}] + \Pr[\up{\tau}^\star \text{ violates precision}].
    \end{equation*}
    In Stage~1, fixed-sequence testing over $\lo{\grid}$ ensures that the probability of rejecting any true recall null hypothesis is bounded by $\delta_R = \delta / 2$. 
    
    In Stage~2, for each fixed $k \in \{1, \dots, \lo{m}\}$, let $E_k$ be the event that the backward fixed-sequence scan over $\up{\grid}$ with fixed lower threshold $\lo{\tau}_k$ commits a Type~I error. 
    By fixed-sequence FWER control, $\Pr[E_k] \le \alpha_P = \delta_P / \lo{m}$ unconditionally. 
    Since the selected threshold satisfies $\lo{\tau}^\star \in \{\lo{\tau}_1, \dots, \lo{\tau}_{\lo{m}}\}$, a Stage~2 error implies that $\bigcup_{k=1}^{\lo{m}} E_k$ occurred. 
    By the union bound:
    \begin{equation*}
        \Pr[\up{\tau}^\star \text{ violates precision}] \le \Pr\Biggl[\bigcup_{k=1}^{\lo{m}} E_k\Biggr] \le \sum_{k=1}^{\lo{m}} \Pr[E_k] \le \lo{m} \alpha_P = \delta_P.
    \end{equation*}
    Combining both stages yields $\text{FWER} \le \delta_R + \delta_P = \delta / 2 + \delta / 2 = \delta$.

    Because $\Call{RecallMet}{}$ and $\Call{PrecisionMet}{}$ are anytime-valid tests (\cref{sec:anytime-tests}), each individual test controls Type~I error uniformly over all stopping times $t \ge 1$, ensuring the FWER bound $\delta$ holds at termination of the repeat--until loop in \cref{alg:tuning}.

\subsection{Warmup Threshold for Adaptive Stopping} \label[appendix]{app:stopping-details}

On highly selective queries, initial tuning batches may contain zero or very few positive records ($Y = 1$), yielding degenerate empirical variance estimates for recall and optimistic thresholds. 
To ensure sufficient positive coverage before permitting $\Call{ShouldStop}{}$ (\cref{sec:stopping-condition}) to halt via $\operatorname{Savings}(B) \le \operatorname{Cost}(B)$ (prior to reaching the sample cap $n_{\max}$), ACTIS sets the minimum positive requirement $n_{\text{min}}$ adaptively from the proxy scores $\mathcal{S}$ and recall error budget $\delta_R = \delta / 2$:
\begin{equation*}
    n_{\text{min}} = \min\Bigl( N, \, \max\bigl(5, \lceil 8 \log(1 / \delta_R) \rceil\bigr), \, \max\bigl(5, \lceil \hat{\pi}_{\text{floor}} N \rceil\bigr) \Bigr),
\end{equation*}
where $\hat{\pi}_{\text{floor}} = \operatorname{clip}(\hat{\pi} / 2, \, 10^{-4}, \, 10^{-2})$ is a conservative lower bound on positive prevalence derived from the mean proxy score $\hat{\pi} = \frac{1}{N}\sum_{x \in \D} \proxy(x)$. 
Here, $\lceil 8 \log(1/\delta_R) \rceil$ corresponds to the sub-Gaussian sample size required to bound the concentration radius $\sqrt{\frac{2\log(1/\delta_R)}{n_+}}$ of the positive sample below $1/2$, $\lceil \hat{\pi}_{\text{floor}} N \rceil$ caps the requirement on small or highly selective workloads, and the floor of $5$ enforces the standard minimum cell count for stable empirical variance estimation.

\subsection{Variance-Minimizing Instrumental Distribution: Derivations and Defensive Mechanisms} \label[appendix]{app:opt-instrumental}

\subsubsection{Variance Objective and Proof of Proposition~\ref{prop:opt-instrumental}}
Evaluating whether candidate thresholds satisfy the recall and precision constraints requires estimating three constituent rates defined in \cref{sec:problem}: the true positive rate $\TPR(\lo{\tau})$, the false positive rate $\FPR(\up{\tau})$, and the positive prevalence $\Prevalence$. 
The importance-weighted estimators of these three constituent rates are:
\begin{align*}
    \widetilde{\TPR}(\lo{\tau}) &= \frac{1}{n} \sum_{i=1}^n w(X_i) Y_i \indicator{\proxy(X_i) \ge \lo{\tau}}, \\
    \widetilde{\FPR}(\up{\tau}) &= \frac{1}{n} \sum_{i=1}^n w(X_i) (1 - Y_i) \indicator{\proxy(X_i) \ge \up{\tau}}, \\
    \widetilde{\Prevalence} &= \frac{1}{n} \sum_{i=1}^n w(X_i) Y_i,
\end{align*}
where $w(X) = p(X) / q(X)$ is the importance weight for record $X$ drawn from instrumental distribution $q$.
To optimize estimation efficiency jointly across the precision and recall constraints, we seek an instrumental distribution $q$ that minimizes the weighted average of the empirical estimation variances across all candidate thresholds in $\lo{\grid}$ and $\up{\grid}$:
\begin{equation}
\begin{split}
    J(q) &\coloneqq \frac{\lambda_{\TPR}}{\lo{m}} \sum_{\lo{\tau} \in \lo{\grid}} \var_q [\widetilde{\TPR}(\lo{\tau})] 
         + \frac{\lambda_{\FPR}}{\up{m}} \sum_{\up{\tau} \in \up{\grid}} \var_q [\widetilde{\FPR}(\up{\tau})] \\
         &\quad + \lambda_{\Prevalence} \var_q [\tilde{\Prevalence}],
\end{split}
\label{eqn:joint_variance_objective}
\end{equation}
where $\lambda_{\TPR}, \lambda_{\FPR}, \lambda_{\Prevalence} > 0$ are weighting coefficients that balance the relative emphasis placed on each constituent rate.

For any estimator of the form $\tilde{\mu} = \frac{1}{n} \sum_{i=1}^n w(X_i) g(X_i, Y_i)$, its variance decomposes as the uncentered second moment minus the squared expectation $\E[g(X, Y)]^2$, which is independent of $q$:
\begin{equation*}
    \var_q[\tilde{\mu}] = \frac{1}{n} \left( \E_{X \sim p, Y}\left[\frac{p(X)}{q(X)} g(X, Y)^2\right] - \E[g(X, Y)]^2 \right).
\end{equation*}
Minimizing $J(q)$ is therefore equivalent to minimizing the weighted sum of uncentered second moments:
\begin{equation*}
    J(q) \propto \E \mathopen{} \left[ \frac{h(\proxy(X), Y)}{q(\proxy(X))} \right],
\end{equation*}
where:
\begin{equation*}
\begin{split}
    h(S, Y) &= \frac{\lambda_{\TPR}}{\lo{m}} \sum_{\lo{\tau} \in \lo{\grid}} \big(Y \indicator{S \ge \lo{\tau}}\big)^2 
        + \frac{\lambda_{\FPR}}{\up{m}} \sum_{\up{\tau} \in \up{\grid}} \big((1 - Y) \indicator{S \ge \up{\tau}}\big)^2 \\
    & \qquad {} + \lambda_{\Prevalence} Y^2.
\end{split}
\end{equation*}
Since the oracle labels and indicators are binary, we have $Y^2 = Y$, $(1 - Y)^2 = 1 - Y$, and $\indicator{\cdot}^2 = \indicator{\cdot}$. 
Averaging across the candidate grids yields:
\begin{equation*}
    h(S, Y) = \lambda_{\TPR} \, Y \frac{\lo{m}(S)}{\lo{m}} 
            + \lambda_{\FPR} \, (1 - Y) \frac{\up{m}(S)}{\up{m}} 
            + \lambda_{\Prevalence} \, Y.
\end{equation*}
Conditioning on the proxy score $S = \proxy(X)$ and invoking the calibration assumption ($\E[Y \mid S] = S$ and $\E[1 - Y \mid S] = 1 - S$), we evaluate the conditional expectation:
\begin{align*}
    c(S) &\coloneqq \E[h(S, Y) \mid \proxy(X) = S] \\
    &= \lambda_{\TPR} \, S \frac{\lo{m}(S)}{\lo{m}} 
     + \lambda_{\FPR} \, (1 - S) \frac{\up{m}(S)}{\up{m}} 
     + \lambda_{\Prevalence} \, S.
\end{align*}

The optimization problem amounts to choosing $q$ that minimizes $\sum_{x \in \D} \frac{c(\proxy(x))}{q(x)}$ subject to $\sum_{x \in \D} q(x) = 1$ with $q(x) > 0$. 
By the Cauchy-Schwarz inequality:
\begin{equation*}
    \left( \sum_{x \in \D} \frac{c(\proxy(x))}{q(x)} \right) \left( \sum_{x \in \D} q(x) \right) \ge \left( \sum_{x \in \D} \sqrt{c(\proxy(x))} \right)^2,
\end{equation*}
with equality holding if and only if $\frac{\sqrt{c(\proxy(x))}}{\sqrt{q(x)}} \propto \sqrt{q(x)}$, which uniquely implies:
\begin{equation*}
    q^*(x) = \frac{\sqrt{c(\proxy(x))}}{\sum_{x' \in \D} \sqrt{c(\proxy(x'))}}.
\end{equation*}

\subsubsection{Derivation of Component Weights} 
To specify the component weights $\lambda_{\TPR}, \lambda_{\FPR}, \lambda_{\Prevalence}$, we standardize the recall and precision constraints by their operational tolerance scales. 
The recall and precision margins can be expressed directly in terms of the component rates:
\begin{align}
    \Delta_R(\lo{\tau}) &= \TPR(\lo{\tau}) - \gamma_R \Prevalence = (1 - \gamma_R) \TPR(\lo{\tau}) - \gamma_R \FNR(\lo{\tau}), \label{eqn:recall-constituent} \\
    \Delta_P(\lo{\tau}, \up{\tau}) &= (1 - \gamma_P) \TPR(\lo{\tau}) - \gamma_P \FPR(\up{\tau}). \label{eqn:precision-constituent}
\end{align}
For recall in~\cref{eqn:recall-constituent}, satisfying target $\gamma_R \approx 1$ depends on detecting rare false negatives; dividing by the slack scale $1 - \gamma_R$ standardizes the true positive term to unit coefficient: $\TPR(\lo{\tau}) - (\frac{\gamma_R}{1 - \gamma_R})\FNR(\lo{\tau}) \ge 0$. 
For precision in~\cref{eqn:precision-constituent}, the risk stems from false positives; dividing by $\gamma_P$ standardizes the false positive term to unit coefficient: $(\frac{1 - \gamma_P}{\gamma_P})\TPR(\lo{\tau}) - \FPR(\up{\tau}) \ge 0$.

The variance of a linear combination of estimators scales quadratically with its linear coefficients. Since every false negative is a positive ($\FNR \le \Prevalence$), we assign the squared false-negative sensitivity to the prevalence term $\tilde{\Prevalence}$ as a threshold-independent surrogate in $J(q)$. 
Reading off the squared sensitivity coefficients across the standardized constraints yields:
\begin{equation*}
    \lambda_{\TPR} = 1 + \left(\frac{1 - \gamma_P}{\gamma_P}\right)^2, \quad
    \lambda_{\FPR} = 1, \quad
    \lambda_{\Prevalence} = \left(\frac{\gamma_R}{1 - \gamma_R}\right)^2.
\end{equation*}
Under these weights, sampling priority concentrates on discovering true positives for recall ($\lambda_{\TPR} \approx 1$) and detecting false positives for precision ($\lambda_{\FPR} = 1$), while scaling up prevalence discovery under tight recall targets ($\gamma_R \to 1$).

\section{Additional Details on the Evaluation} \label[appendix]{app:evaluation}

This appendix provides comprehensive dataset statistics and complete empirical results complementing \cref{sec:evaluation}.

\myparagraph{Benchmark Statistics.}
\cref{tab:dataset-stats} summarizes all 42 benchmarks evaluated in our study, spanning 23 workloads across pre-LLM computer vision and NLP tasks~\citep{kang2020approximate}, general text corpora~\citep{zeighami2025cut}, and technical documents~\citep{zhang2026scaledoc}. 
Alongside workload size $N$, filtered record counts, and ground-truth positive prevalence $\pi$, we characterize proxy quality using Precision-Recall AUC (PR-AUC), calibration bias ($\E[\proxy(X) - Y]$), Expected Calibration Error (ECE, computed over 10 equal-width bins on $[0, 1]$), and score saturation rate (the fraction of records assigned exact scores of $0$ or $1$). 
Note that while cloud-hosted GPT-5.6-luna achieves higher PR-AUC than local Llama-3.2-1B across all text workloads, it frequently produces saturated confidence scores (up to $55.8\%$ of records), motivating the tail-tolerance clipping in \cref{sec:instrumental-design}.

\myparagraph{Soundness Audit.}
\cref{tab:soundness_disaggregated_joint} disaggregates the statistical soundness audit from \cref{sec:exp-soundness} (\cref{tab:soundness_summary_joint}) across individual error budgets $\delta \in \{0.01, 0.05, 0.10\}$ and accuracy targets $\gamma \in \{0.70, 0.80, 0.90, 0.95\}$. 
Across all twelve $(\delta, \gamma)$ configurations, both ACTIS and ACTIS-U achieve $0.0\%$ certified unsoundness, whereas LOTUS and ScaleDoc exhibit high failure rates that worsen as $\gamma$ increases. 
The ablation of ACTIS without runtime safeguards incurs unsoundness on $2.4\%$--$7.1\%$ of benchmarks across all regimes, confirming that our finite-sample safeguards (\cref{sec:asymptotic-safeguards}) are essential for maintaining strict Type~I error control under asymptotic testing.

\myparagraph{Complete Performance Results.}
\cref{tab:all_benchmarks_performance_mean_only} reports the mean oracle call rate (OCR) and CPU tuning latency across all 42 benchmarks at $\gamma = 0.95$ and $\delta = 0.05$, expanding upon the representative subset in \cref{tab:representative_performance}.

\begin{table*}
  \centering
  \caption{Summary statistics across all 42 benchmarks: workload size ($N$), records excluded by API content filters (Missing), ground-truth positive prevalence (Pos.\ rate), area under the proxy precision-recall curve (PR-AUC), mean calibration bias $\E[\proxy(X) - Y]$ (Cal.\ Bias), expected calibration error over 10 equal-width bins (ECE), and fraction of records with saturated proxy scores in $\{0, 1\}$ (Sat.\ Rate).
  }
  \label{tab:dataset-stats}
  \footnotesize
  \begin{tabular}{lrllrrrrrr}
    \toprule
    \textbf{Dataset} & \textbf{$N$} & \textbf{Query} & \textbf{Missing} & \textbf{Pos.\ rate $\Prevalence$} & \textbf{Proxy Model} & \textbf{PR-AUC} & \textbf{Cal.\ Bias} & \textbf{ECE} & \textbf{Sat.\ Rate} \\
    \midrule
    Onto & 11,165 & --- & 0 & 0.025 & UFT~\citep{choi2018ultra} & 0.830 & +0.008 & 0.009 & 0.000 \\
    \midrule[0.2pt]
    ImageNet & 50,000 & --- & 0 & 0.001 & ResNet-50~\citep{he2016deep} & 0.982 & +0.000 & 0.000 & 0.000 \\
    \midrule[0.2pt]
    Jackson & 973,085 & --- & 0 & 0.292 & ResNet-50~\citep{he2016deep} & 0.880 & +0.017 & 0.023 & 0.007 \\
    \midrule[0.2pt]
    TACRED & 22,631 & --- & 0 & 0.024 & SpanBERT~\citep{joshi2020spanbert} & 0.794 & +0.001 & 0.007 & 0.000 \\
    \midrule[0.2pt]
    Court & 1,000 & --- & 4 & 0.650 & GPT-5.6-luna & 0.897 & -0.057 & 0.201 & 0.255 \\
    \cmidrule[0.2pt]{6-10}
    & & & & & Llama-3.2-1B & 0.645 & +0.292 & 0.292 & 0.000 \\
    \midrule[0.2pt]
    Review & 10,000 & --- & 8 & 0.059 & GPT-5.6-luna & 0.823 & +0.035 & 0.042 & 0.509 \\
    \cmidrule[0.2pt]{6-10}
    & & & & & Llama-3.2-1B & 0.124 & +0.618 & 0.618 & 0.000 \\
    \midrule[0.2pt]
    Screenplay & 2,858 & --- & 4 & 0.865 & GPT-5.6-luna & 0.965 & +0.058 & 0.086 & 0.082 \\
    \cmidrule[0.2pt]{6-10}
    & & & & & Llama-3.2-1B & 0.886 & +0.085 & 0.104 & 0.000 \\
    \midrule[0.2pt]
    Wiki & 10,000 & --- & 1 & 0.342 & GPT-5.6-luna & 0.868 & -0.166 & 0.173 & 0.243 \\
    \cmidrule[0.2pt]{6-10}
    & & & & & Llama-3.2-1B & 0.362 & +0.289 & 0.289 & 0.000 \\
    \midrule[0.2pt]
    PubMed & 10,000 & 0\textsubscript{ext} & 0 & 0.006 & GPT-5.6-luna & 0.024 & -0.005 & 0.006 & 0.173 \\
    \cmidrule[0.2pt]{6-10}
    & & & & & Llama-3.2-1B & 0.004 & +0.503 & 0.503 & 0.000 \\
    \cmidrule[0.2pt]{3-10}
    & & 1\textsubscript{ext} & 0 & 0.615 & GPT-5.6-luna & 0.999 & +0.055 & 0.055 & 0.416 \\
    \cmidrule[0.2pt]{6-10}
    & & & & & Llama-3.2-1B & 0.867 & +0.032 & 0.138 & 0.000 \\
    \cmidrule[0.2pt]{3-10}
    & & 2\textsubscript{ext} & 0 & 0.584 & GPT-5.6-luna & 0.996 & -0.012 & 0.020 & 0.160 \\
    \cmidrule[0.2pt]{6-10}
    & & & & & Llama-3.2-1B & 0.731 & -0.177 & 0.177 & 0.000 \\
    \cmidrule[0.2pt]{3-10}
    & & 0 & 0 & 0.790 & GPT-5.6-luna & 0.987 & +0.043 & 0.060 & 0.460 \\
    \cmidrule[0.2pt]{6-10}
    & & & & & Llama-3.2-1B & 0.932 & -0.285 & 0.285 & 0.000 \\
    \cmidrule[0.2pt]{3-10}
    & & 4 & 0 & 0.680 & GPT-5.6-luna & 0.991 & +0.054 & 0.065 & 0.422 \\
    \cmidrule[0.2pt]{6-10}
    & & & & & Llama-3.2-1B & 0.884 & -0.022 & 0.066 & 0.000 \\
    \cmidrule[0.2pt]{3-10}
    & & 10 & 0 & 0.122 & GPT-5.6-luna & 0.995 & +0.010 & 0.011 & 0.558 \\
    \cmidrule[0.2pt]{6-10}
    & & & & & Llama-3.2-1B & 0.895 & +0.084 & 0.117 & 0.000 \\
    \midrule[0.2pt]
    BigPatent & 10,000 & 0\textsubscript{ext} & 0 & 0.121 & GPT-5.6-luna & 0.988 & +0.007 & 0.012 & 0.512 \\
    \cmidrule[0.2pt]{6-10}
    & & & & & Llama-3.2-1B & 0.714 & +0.019 & 0.091 & 0.000 \\
    \cmidrule[0.2pt]{3-10}
    & & 1\textsubscript{ext} & 0 & 0.676 & GPT-5.6-luna & 0.966 & -0.243 & 0.243 & 0.061 \\
    \cmidrule[0.2pt]{6-10}
    & & & & & Llama-3.2-1B & 0.723 & -0.366 & 0.366 & 0.000 \\
    \cmidrule[0.2pt]{3-10}
    & & 2\textsubscript{ext} & 0 & 0.379 & GPT-5.6-luna & 0.952 & -0.030 & 0.080 & 0.307 \\
    \cmidrule[0.2pt]{6-10}
    & & & & & Llama-3.2-1B & 0.575 & -0.196 & 0.196 & 0.000 \\
    \cmidrule[0.2pt]{3-10}
    & & 0 & 0 & 0.231 & GPT-5.6-luna & 0.986 & +0.022 & 0.028 & 0.477 \\
    \cmidrule[0.2pt]{6-10}
    & & & & & Llama-3.2-1B & 0.840 & -0.126 & 0.126 & 0.000 \\
    \cmidrule[0.2pt]{3-10}
    & & 4 & 0 & 0.061 & GPT-5.6-luna & 0.214 & -0.042 & 0.058 & 0.442 \\
    \cmidrule[0.2pt]{6-10}
    & & & & & Llama-3.2-1B & 0.052 & +0.069 & 0.085 & 0.000 \\
    \cmidrule[0.2pt]{3-10}
    & & 10 & 0 & 0.056 & GPT-5.6-luna & 0.936 & -0.013 & 0.014 & 0.556 \\
    \cmidrule[0.2pt]{6-10}
    & & & & & Llama-3.2-1B & 0.282 & +0.088 & 0.088 & 0.000 \\
    \midrule[0.2pt]
    GovReport & 10,000 & 0 & 2 & 0.207 & GPT-5.6-luna & 0.982 & +0.023 & 0.027 & 0.492 \\
    \cmidrule[0.2pt]{6-10}
    & & & & & Llama-3.2-1B & 0.723 & +0.114 & 0.136 & 0.000 \\
    \cmidrule[0.2pt]{3-10}
    & & 4 & 2 & 0.199 & GPT-5.6-luna & 0.837 & +0.116 & 0.136 & 0.156 \\
    \cmidrule[0.2pt]{6-10}
    & & & & & Llama-3.2-1B & 0.375 & +0.281 & 0.281 & 0.000 \\
    \cmidrule[0.2pt]{3-10}
    & & 10 & 2 & 0.146 & GPT-5.6-luna & 0.905 & -0.009 & 0.040 & 0.437 \\
    \cmidrule[0.2pt]{6-10}
    & & & & & Llama-3.2-1B & 0.575 & +0.198 & 0.198 & 0.000 \\
    \bottomrule
  \end{tabular}
\end{table*}

\begin{table*}[p]
\centering
\caption{Statistical soundness across 42 benchmarks disaggregated by error budget $\delta$ and accuracy target $\gamma$. 
Using 95\% Clopper-Pearson confidence intervals $[\tilde{\delta}_\text{LB}, \tilde{\delta}_\text{UB}]$ over 1,000 trials, we report the percentage of benchmarks certified \emph{sound} ($\tilde{\delta}_\text{UB} < \delta$), \emph{inconclusive} ($\delta \in [\tilde{\delta}_\text{LB}, \tilde{\delta}_\text{UB}]$), or \emph{unsound} (in \badvalue{red}; $\tilde{\delta}_\text{LB} > \delta$). 
$\tilde{\delta}_{\text{unsound}}$ and $\tilde{\delta}_{\max}$ denote the mean empirical failure rate over certified unsound benchmarks and the maximum empirical failure rate across all benchmarks, respectively.}
\addvspace{-6pt}
\renewcommand{\arraystretch}{0.85}
\label{tab:soundness_disaggregated_joint}
\footnotesize
\setlength{\tabcolsep}{3.5pt}
\begin{tabular}{lll rrr rr}
\toprule
& & & \multicolumn{3}{c}{\textbf{Percentage Certified}} & & \\
\cmidrule(lr){4-6}
$\delta$ & $\gamma$ & \textbf{Method} & \textbf{Sound} & \textbf{Inconcl.} & \textbf{Unsound} & $\tilde{\delta}_{\text{unsound}}$ & $\tilde{\delta}_{\max}$ \\
\midrule
0.01 & 0.70 & ACTIS & 100.0 & 0.0 & 0.0 & \text{---} & 0.000 \\
 &  & ACTIS (w/o safeguards) & 97.6 & 0.0 & \badvalue{2.4} & 0.069 & 0.069 \\
 &  & ACTIS-U & 100.0 & 0.0 & 0.0 & \text{---} & 0.001 \\
 &  & BARGAIN & 85.7 & 14.3 & 0.0 & \text{---} & 0.005 \\
 &  & ScaleDoc & 69.0 & 4.8 & \badvalue{26.2} & 0.345 & 1.000 \\
 &  & LOTUS & 28.6 & 0.0 & \badvalue{71.4} & 0.792 & 1.000 \\
\cmidrule{2-8}
 & 0.80 & ACTIS & 100.0 & 0.0 & 0.0 & \text{---} & 0.003 \\
 &  & ACTIS (w/o safeguards) & 97.6 & 0.0 & \badvalue{2.4} & 0.129 & 0.129 \\
 &  & ACTIS-U & 100.0 & 0.0 & 0.0 & \text{---} & 0.001 \\
 &  & BARGAIN & 88.1 & 11.9 & 0.0 & \text{---} & 0.006 \\
 &  & ScaleDoc & 54.8 & 4.8 & \badvalue{40.5} & 0.329 & 0.999 \\
 &  & LOTUS & 21.4 & 0.0 & \badvalue{78.6} & 0.869 & 1.000 \\
\cmidrule{2-8}
 & 0.90 & ACTIS & 100.0 & 0.0 & 0.0 & \text{---} & 0.001 \\
 &  & ACTIS (w/o safeguards) & 95.2 & 0.0 & \badvalue{4.8} & 0.083 & 0.146 \\
 &  & ACTIS-U & 100.0 & 0.0 & 0.0 & \text{---} & 0.000 \\
 &  & BARGAIN & 90.5 & 9.5 & 0.0 & \text{---} & 0.006 \\
 &  & ScaleDoc & 31.0 & 2.4 & \badvalue{66.7} & 0.336 & 1.000 \\
 &  & LOTUS & 7.1 & 7.1 & \badvalue{85.7} & 0.866 & 1.000 \\
\cmidrule{2-8}
 & 0.95 & ACTIS & 97.6 & 2.4 & 0.0 & \text{---} & 0.004 \\
 &  & ACTIS (w/o safeguards) & 85.7 & 9.5 & \badvalue{4.8} & 0.157 & 0.217 \\
 &  & ACTIS-U & 100.0 & 0.0 & 0.0 & \text{---} & 0.000 \\
 &  & BARGAIN & 100.0 & 0.0 & 0.0 & \text{---} & 0.003 \\
 &  & ScaleDoc & 21.4 & 0.0 & \badvalue{78.6} & 0.317 & 0.934 \\
 &  & LOTUS & 4.8 & 4.8 & \badvalue{90.5} & 0.841 & 1.000 \\
\midrule
0.05 & 0.70 & ACTIS & 100.0 & 0.0 & 0.0 & \text{---} & 0.030 \\
 &  & ACTIS (w/o safeguards) & 97.6 & 0.0 & \badvalue{2.4} & 0.102 & 0.102 \\
 &  & ACTIS-U & 100.0 & 0.0 & 0.0 & \text{---} & 0.002 \\
 &  & BARGAIN & 100.0 & 0.0 & 0.0 & \text{---} & 0.020 \\
 &  & ScaleDoc & 76.2 & 7.1 & \badvalue{16.7} & 0.516 & 1.000 \\
 &  & LOTUS & 28.6 & 0.0 & \badvalue{71.4} & 0.810 & 1.000 \\
\cmidrule{2-8}
 & 0.80 & ACTIS & 97.6 & 2.4 & 0.0 & \text{---} & 0.043 \\
 &  & ACTIS (w/o safeguards) & 97.6 & 0.0 & \badvalue{2.4} & 0.160 & 0.160 \\
 &  & ACTIS-U & 100.0 & 0.0 & 0.0 & \text{---} & 0.001 \\
 &  & BARGAIN & 100.0 & 0.0 & 0.0 & \text{---} & 0.027 \\
 &  & ScaleDoc & 61.9 & 9.5 & \badvalue{28.6} & 0.450 & 0.999 \\
 &  & LOTUS & 21.4 & 2.4 & \badvalue{76.2} & 0.904 & 1.000 \\
\cmidrule{2-8}
 & 0.90 & ACTIS & 100.0 & 0.0 & 0.0 & \text{---} & 0.007 \\
 &  & ACTIS (w/o safeguards) & 95.2 & 2.4 & \badvalue{2.4} & 0.173 & 0.173 \\
 &  & ACTIS-U & 100.0 & 0.0 & 0.0 & \text{---} & 0.002 \\
 &  & BARGAIN & 100.0 & 0.0 & 0.0 & \text{---} & 0.030 \\
 &  & ScaleDoc & 42.9 & 7.1 & \badvalue{50.0} & 0.436 & 1.000 \\
 &  & LOTUS & 16.7 & 0.0 & \badvalue{83.3} & 0.908 & 1.000 \\
\cmidrule{2-8}
 & 0.95 & ACTIS & 100.0 & 0.0 & 0.0 & \text{---} & 0.007 \\
 &  & ACTIS (w/o safeguards) & 92.9 & 0.0 & \badvalue{7.1} & 0.174 & 0.232 \\
 &  & ACTIS-U & 100.0 & 0.0 & 0.0 & \text{---} & 0.002 \\
 &  & BARGAIN & 100.0 & 0.0 & 0.0 & \text{---} & 0.022 \\
 &  & ScaleDoc & 26.2 & 9.5 & \badvalue{64.3} & 0.376 & 0.934 \\
 &  & LOTUS & 7.1 & 2.4 & \badvalue{90.5} & 0.862 & 1.000 \\
\midrule
0.10 & 0.70 & ACTIS & 100.0 & 0.0 & 0.0 & \text{---} & 0.061 \\
 &  & ACTIS (w/o safeguards) & 97.6 & 0.0 & \badvalue{2.4} & 0.125 & 0.125 \\
 &  & ACTIS-U & 100.0 & 0.0 & 0.0 & \text{---} & 0.004 \\
 &  & BARGAIN & 100.0 & 0.0 & 0.0 & \text{---} & 0.039 \\
 &  & ScaleDoc & 85.7 & 0.0 & \badvalue{14.3} & 0.591 & 1.000 \\
 &  & LOTUS & 28.6 & 4.8 & \badvalue{66.7} & 0.871 & 1.000 \\
\cmidrule{2-8}
 & 0.80 & ACTIS & 100.0 & 0.0 & 0.0 & \text{---} & 0.065 \\
 &  & ACTIS (w/o safeguards) & 97.6 & 0.0 & \badvalue{2.4} & 0.186 & 0.186 \\
 &  & ACTIS-U & 100.0 & 0.0 & 0.0 & \text{---} & 0.002 \\
 &  & BARGAIN & 100.0 & 0.0 & 0.0 & \text{---} & 0.048 \\
 &  & ScaleDoc & 73.8 & 2.4 & \badvalue{23.8} & 0.522 & 0.999 \\
 &  & LOTUS & 23.8 & 0.0 & \badvalue{76.2} & 0.909 & 1.000 \\
\cmidrule{2-8}
 & 0.90 & ACTIS & 100.0 & 0.0 & 0.0 & \text{---} & 0.041 \\
 &  & ACTIS (w/o safeguards) & 97.6 & 0.0 & \badvalue{2.4} & 0.190 & 0.190 \\
 &  & ACTIS-U & 100.0 & 0.0 & 0.0 & \text{---} & 0.003 \\
 &  & BARGAIN & 100.0 & 0.0 & 0.0 & \text{---} & 0.049 \\
 &  & ScaleDoc & 52.4 & 4.8 & \badvalue{42.9} & 0.492 & 1.000 \\
 &  & LOTUS & 16.7 & 0.0 & \badvalue{83.3} & 0.916 & 1.000 \\
\cmidrule{2-8}
 & 0.95 & ACTIS & 100.0 & 0.0 & 0.0 & \text{---} & 0.009 \\
 &  & ACTIS (w/o safeguards) & 92.9 & 2.4 & \badvalue{4.8} & 0.264 & 0.283 \\
 &  & ACTIS-U & 100.0 & 0.0 & 0.0 & \text{---} & 0.004 \\
 &  & BARGAIN & 100.0 & 0.0 & 0.0 & \text{---} & 0.039 \\
 &  & ScaleDoc & 38.1 & 7.1 & \badvalue{54.8} & 0.426 & 0.934 \\
 &  & LOTUS & 7.1 & 2.4 & \badvalue{90.5} & 0.874 & 1.000 \\
\bottomrule
\end{tabular}
\end{table*}

\begin{table*}[p]
\centering
\caption{Oracle call rate (\%) and tuner latency (ms) across all benchmarks for precision\slash recall target $\gamma = 0.95$ and error budget $\delta = 0.05$. \badvalue{Red*} indicates certified unsoundness; boldface highlights the best result among statistically sound methods.}
\label{tab:all_benchmarks_performance_mean_only}
\label{tab:oracle_call_rates_all_benchmarks_mean_only}
\label{tab:tuning_latency_all_benchmarks_mean_only}
\footnotesize
\setlength{\tabcolsep}{4.2pt}
\begin{tabular}{llr rrrrrr rrrrr}
\toprule
\multicolumn{2}{c}{\textbf{Benchmark}} & & \multicolumn{6}{c}{\textbf{Oracle Call Rate (\%)}} & \multicolumn{5}{c}{\textbf{Tuner Latency (ms)}} \\
\cmidrule(lr){1-2} \cmidrule(lr){4-9} \cmidrule(lr){10-14}
\textbf{Workload} & \textbf{Proxy} & $\mathbf{N}$ & \textbf{Ideal} & \textbf{ACTIS} & \textbf{ACTIS-U} & \textbf{BARGAIN} & \textbf{ScaleDoc} & \textbf{LOTUS} & \textbf{ACTIS} & \textbf{ACTIS-U} & \textbf{BARGAIN} & \textbf{ScaleDoc} & \textbf{LOTUS} \\
\midrule
Onto & UFT & 11k & 6.2 & \textbf{32.1} & 61.3 & 62.6 & \badvalue{28.4\rlap{*}} & \badvalue{63.2\rlap{*}} & 17.2 & 42.6 & \textbf{11.2} & \badvalue{4.4\rlap{*}} & \badvalue{1.6\rlap{*}} \\
\addlinespace
ImageNet & ResNet-50 & 50k & 0.0 & 19.6 & 100.0 & 100.0 & 10.1 & \textbf{5.1} & 48.1 & 191.6 & 123.2 & 7.1 & \textbf{6.2} \\
\addlinespace
Jackson & ResNet-50 & 973k & 35.4 & \textbf{45.0} & 46.4 & 81.5 & 58.7 & \badvalue{25.7\rlap{*}} & 3854.3 & 10110.6 & 3580.0 & \textbf{109.6} & \badvalue{171.3\rlap{*}} \\
\addlinespace
TACRED & SpanBERT & 23k & 6.5 & \textbf{32.1} & 54.4 & 51.3 & \badvalue{43.5\rlap{*}} & \badvalue{35.9\rlap{*}} & 43.7 & 69.1 & \textbf{22.2} & \badvalue{5.5\rlap{*}} & \badvalue{3.0\rlap{*}} \\
\addlinespace
Court & GPT-5.6-luna & 1k & 44.5 & \textbf{80.2} & 88.1 & 94.0 & \badvalue{60.8\rlap{*}} & \badvalue{22.3\rlap{*}} & 3.6 & 5.6 & \textbf{2.2} & \badvalue{3.3\rlap{*}} & \badvalue{0.6\rlap{*}} \\
 & Llama-3.2-1B & 1k & 85.2 & \textbf{96.8} & 97.6 & 98.6 & \badvalue{95.4\rlap{*}} & \badvalue{67.3\rlap{*}} & 6.1 & 8.0 & \textbf{2.6} & \badvalue{4.3\rlap{*}} & \badvalue{0.7\rlap{*}} \\
\addlinespace
Review & GPT-5.6-luna & 10k & 10.2 & \textbf{32.8} & 51.7 & 65.3 & \badvalue{22.8\rlap{*}} & \badvalue{7.4\rlap{*}} & \textbf{10.7} & 18.4 & 12.3 & \badvalue{4.3\rlap{*}} & \badvalue{1.6\rlap{*}} \\
 & Llama-3.2-1B & 10k & 89.0 & 98.1 & \textbf{96.9} & 97.9 & \badvalue{94.6\rlap{*}} & \badvalue{77.2\rlap{*}} & 25.4 & 47.0 & \textbf{17.1} & \badvalue{4.7\rlap{*}} & \badvalue{1.6\rlap{*}} \\
\addlinespace
Screenplay & GPT-5.6-luna & 3k & 15.3 & \textbf{49.7} & 53.3 & 93.4 & \badvalue{73.0\rlap{*}} & \badvalue{34.0\rlap{*}} & 13.8 & 23.1 & \textbf{6.5} & \badvalue{5.1\rlap{*}} & \badvalue{1.1\rlap{*}} \\
 & Llama-3.2-1B & 3k & 53.9 & \textbf{84.1} & 85.5 & 97.5 & 94.8 & \badvalue{52.0\rlap{*}} & 44.2 & 87.2 & 4.1 & \textbf{3.9} & \badvalue{0.9\rlap{*}} \\
\addlinespace
Wiki & GPT-5.6-luna & 10k & 37.0 & 71.1 & \textbf{68.6} & 83.1 & \badvalue{82.3\rlap{*}} & \badvalue{26.1\rlap{*}} & 69.5 & 129.1 & \textbf{13.7} & \badvalue{3.4\rlap{*}} & \badvalue{1.6\rlap{*}} \\
 & Llama-3.2-1B & 10k & 91.7 & \textbf{97.0} & 97.2 & 98.0 & \badvalue{95.2\rlap{*}} & \badvalue{55.6\rlap{*}} & 45.5 & 88.3 & \textbf{20.5} & \badvalue{5.4\rlap{*}} & \badvalue{2.4\rlap{*}} \\
\addlinespace
PubMed Q0 & GPT-5.6-luna & 10k & 0.0 & \textbf{15.3} & 15.8 & 24.8 & 42.3 & \badvalue{9.4\rlap{*}} & \textbf{3.9} & 4.5 & 5.2 & 6.2 & \badvalue{1.8\rlap{*}} \\
 & Llama-3.2-1B & 10k & 37.7 & 61.6 & 62.1 & 92.1 & \textbf{56.1} & \badvalue{48.8\rlap{*}} & 85.6 & 214.4 & 27.2 & \textbf{5.0} & \badvalue{1.7\rlap{*}} \\
\addlinespace
PubMed Q0\textsubscript{ext} & GPT-5.6-luna & 10k & 100.0 & 100.0 & 100.0 & 100.0 & \textbf{94.9} & \badvalue{55.9\rlap{*}} & 14.1 & 36.2 & 20.3 & \textbf{4.1} & \badvalue{1.8\rlap{*}} \\
 & Llama-3.2-1B & 10k & 97.9 & \textbf{100.0} & \textbf{100.0} & \textbf{100.0} & \badvalue{98.5\rlap{*}} & \badvalue{89.5\rlap{*}} & 21.0 & 48.3 & \textbf{19.8} & \badvalue{5.0\rlap{*}} & \badvalue{1.7\rlap{*}} \\
\addlinespace
PubMed Q1\textsubscript{ext} & GPT-5.6-luna & 10k & 0.0 & 11.4 & 10.0 & 16.1 & 29.7 & \textbf{9.4} & 3.9 & 3.9 & 2.6 & 5.1 & \textbf{1.7} \\
 & Llama-3.2-1B & 10k & 50.1 & 71.9 & \textbf{71.8} & 85.1 & 73.8 & \badvalue{67.4\rlap{*}} & 77.2 & 174.7 & 13.3 & \textbf{5.2} & \badvalue{1.8\rlap{*}} \\
\addlinespace
PubMed Q2\textsubscript{ext} & GPT-5.6-luna & 10k & 0.0 & 10.7 & 11.1 & 16.5 & \textbf{5.6} & \badvalue{9.3\rlap{*}} & 7.1 & 12.8 & \textbf{3.1} & 6.5 & \badvalue{1.8\rlap{*}} \\
 & Llama-3.2-1B & 10k & 76.5 & \textbf{89.2} & 89.8 & 95.7 & 89.6 & \badvalue{57.1\rlap{*}} & 67.9 & 150.5 & 19.6 & \textbf{5.2} & \badvalue{1.9\rlap{*}} \\
\addlinespace
PubMed Q4 & GPT-5.6-luna & 10k & 0.0 & \textbf{15.2} & 15.4 & 23.1 & 38.9 & \badvalue{9.4\rlap{*}} & \textbf{4.2} & 5.1 & 4.9 & 5.8 & \badvalue{1.8\rlap{*}} \\
 & Llama-3.2-1B & 10k & 51.1 & 71.8 & 71.8 & 90.6 & \textbf{70.9} & \badvalue{61.3\rlap{*}} & 83.6 & 188.2 & 19.9 & \textbf{5.2} & \badvalue{1.8\rlap{*}} \\
\addlinespace
PubMed Q10 & GPT-5.6-luna & 10k & 0.0 & \textbf{11.6} & 24.8 & 19.6 & \badvalue{9.7\rlap{*}} & \badvalue{7.5\rlap{*}} & 3.5 & 3.5 & \textbf{2.7} & \badvalue{4.8\rlap{*}} & \badvalue{1.6\rlap{*}} \\
 & Llama-3.2-1B & 10k & 13.2 & 57.3 & \textbf{50.1} & 64.6 & \badvalue{36.0\rlap{*}} & 61.1 & 68.0 & 37.2 & 12.1 & \badvalue{5.5\rlap{*}} & \textbf{1.7} \\
\addlinespace
BigPatent Q0 & GPT-5.6-luna & 10k & 1.8 & \textbf{15.6} & 21.8 & 22.6 & 20.6 & \badvalue{8.6\rlap{*}} & \textbf{3.7} & 4.2 & 5.3 & 6.4 & \badvalue{2.3\rlap{*}} \\
 & Llama-3.2-1B & 10k & 30.1 & 52.8 & 53.5 & 64.0 & 61.6 & \textbf{46.2} & 36.4 & 56.2 & 14.5 & 5.3 & \textbf{1.9} \\
\addlinespace
BigPatent Q0\textsubscript{ext} & GPT-5.6-luna & 10k & 0.7 & \textbf{14.3} & 26.5 & 21.2 & \badvalue{11.7\rlap{*}} & \badvalue{7.6\rlap{*}} & \textbf{3.4} & 3.8 & 4.9 & \badvalue{6.2\rlap{*}} & \badvalue{2.4\rlap{*}} \\
 & Llama-3.2-1B & 10k & 60.2 & 85.9 & \textbf{84.8} & 94.0 & 86.4 & \badvalue{71.6\rlap{*}} & 69.3 & 113.9 & 18.9 & \textbf{5.4} & \badvalue{1.9\rlap{*}} \\
\addlinespace
BigPatent Q1\textsubscript{ext} & GPT-5.6-luna & 10k & 18.5 & 43.6 & \textbf{41.0} & 56.8 & 52.4 & \badvalue{9.2\rlap{*}} & 43.5 & 79.2 & 14.7 & \textbf{4.7} & \badvalue{2.6\rlap{*}} \\
 & Llama-3.2-1B & 10k & 80.3 & 90.8 & 91.2 & 98.0 & \textbf{89.8} & \badvalue{39.9\rlap{*}} & 72.1 & 161.2 & 20.8 & \textbf{5.5} & \badvalue{1.9\rlap{*}} \\
\addlinespace
BigPatent Q2\textsubscript{ext} & GPT-5.6-luna & 10k & 17.0 & \textbf{36.7} & 38.4 & 44.9 & \badvalue{65.0\rlap{*}} & \badvalue{9.0\rlap{*}} & 11.3 & 19.4 & \textbf{10.5} & \badvalue{5.2\rlap{*}} & \badvalue{2.6\rlap{*}} \\
 & Llama-3.2-1B & 10k & 80.9 & \textbf{90.9} & 91.4 & 94.5 & 92.7 & \badvalue{54.3\rlap{*}} & 56.0 & 112.8 & 19.7 & \textbf{5.2} & \badvalue{1.9\rlap{*}} \\
\addlinespace
BigPatent Q4 & GPT-5.6-luna & 10k & 99.1 & 100.0 & 99.7 & \textbf{99.4} & \badvalue{98.9\rlap{*}} & \badvalue{79.1\rlap{*}} & \textbf{21.2} & 34.3 & 24.3 & \badvalue{5.1\rlap{*}} & \badvalue{2.5\rlap{*}} \\
 & Llama-3.2-1B & 10k & 95.2 & 99.1 & 99.1 & \textbf{99.0} & \badvalue{98.6\rlap{*}} & \badvalue{72.0\rlap{*}} & 29.2 & 50.8 & \textbf{20.4} & \badvalue{5.3\rlap{*}} & \badvalue{1.8\rlap{*}} \\
\addlinespace
BigPatent Q10 & GPT-5.6-luna & 10k & 2.6 & \textbf{53.0} & 66.1 & 78.3 & \badvalue{85.1\rlap{*}} & \badvalue{5.3\rlap{*}} & \textbf{17.6} & 26.5 & 19.2 & \badvalue{5.3\rlap{*}} & \badvalue{2.7\rlap{*}} \\
 & Llama-3.2-1B & 10k & 68.8 & 90.7 & \textbf{90.3} & 92.2 & \badvalue{85.7\rlap{*}} & \badvalue{83.2\rlap{*}} & 38.1 & 55.7 & \textbf{19.1} & \badvalue{5.6\rlap{*}} & \badvalue{1.8\rlap{*}} \\
\addlinespace
GovReport Q0 & GPT-5.6-luna & 10k & 2.0 & \textbf{15.6} & 23.2 & 21.3 & 18.9 & \badvalue{8.5\rlap{*}} & 4.9 & 6.9 & \textbf{4.5} & 4.9 & \badvalue{1.8\rlap{*}} \\
 & Llama-3.2-1B & 10k & 65.9 & \textbf{84.0} & 84.3 & 90.3 & \badvalue{79.0\rlap{*}} & \badvalue{54.8\rlap{*}} & 70.7 & 117.1 & \textbf{17.4} & \badvalue{5.4\rlap{*}} & \badvalue{2.0\rlap{*}} \\
\addlinespace
GovReport Q4 & GPT-5.6-luna & 10k & 28.1 & \textbf{49.0} & 53.5 & 62.5 & \badvalue{41.2\rlap{*}} & \badvalue{15.7\rlap{*}} & 14.8 & 21.8 & \textbf{12.0} & \badvalue{4.8\rlap{*}} & \badvalue{1.9\rlap{*}} \\
 & Llama-3.2-1B & 10k & 83.3 & 93.1 & 93.3 & 95.7 & \textbf{89.8} & \badvalue{63.0\rlap{*}} & 50.4 & 78.5 & 18.2 & \textbf{5.3} & \badvalue{2.0\rlap{*}} \\
\addlinespace
GovReport Q10 & GPT-5.6-luna & 10k & 12.5 & \textbf{35.6} & 43.1 & 52.7 & \badvalue{77.4\rlap{*}} & \badvalue{7.9\rlap{*}} & 13.3 & 17.9 & \textbf{10.2} & \badvalue{4.6\rlap{*}} & \badvalue{1.9\rlap{*}} \\
 & Llama-3.2-1B & 10k & 62.6 & 84.1 & \textbf{83.9} & 86.9 & \badvalue{75.4\rlap{*}} & \badvalue{62.8\rlap{*}} & 58.3 & 81.5 & \textbf{16.7} & \badvalue{5.6\rlap{*}} & \badvalue{2.0\rlap{*}} \\
\bottomrule
\end{tabular}
\end{table*}

\section{Statistical Soundness of Prior Work} \label[appendix]{app:soundness-analysis}

In this appendix, we briefly discuss potential statistical and algorithmic issues in existing semantic filter approximation methods that claim probabilistic accuracy guarantees: LOTUS~\citep{patel2025semantic}, SUPG-IT~\citep{liskowski2026streaming}, ScaleDoc~\citep{zhang2026scaledoc}, Stretto~\citep{sanmartino2026stretto}, and Clustering-Sampling-Voting (CSV)~\citep{hou2026linear}. 
We emphasize that this discussion is non-exhaustive and reflects issues identified from a reading of the respective papers and, for the baselines evaluated in \cref{sec:evaluation} (LOTUS and ScaleDoc), a cursory inspection of their released code. 
We have not attempted to catalog every possible issue or empirically isolate the causal impact of each individual factor on the failure rates observed in \cref{sec:exp-soundness}.

\subsection{LOTUS}
LOTUS~\citep{patel2025semantic} extends SUPG~\citep{kang2020approximate} to tune a two-threshold cascade for simultaneous precision and recall targets using a single importance-sampled pilot set of size $s$. 
While \citet{patel2025semantic} state that the joint guarantee is achieved by splitting the failure probability ($\delta / 2$ per metric) and applying normal confidence bounds with a Bonferroni correction over $s$, the paper does not provide a formal theorem or proof. 
In their tuning procedure and released implementation, several issues can undermine coverage:
(i)~thresholds are selected using asymptotic Gaussian confidence bounds with plug-in empirical Bernoulli standard deviations $\hat{\sigma}_k = \sqrt{\hat{\mu}_k(1 - \hat{\mu}_k)}$, which collapse to zero ($\hat{\sigma}_k = 0$) whenever the top-$k$ highest-scoring tuning records are all positive ($\hat{\mu}_k = 1$), yielding a zero-width confidence interval that can select an overly aggressive upper threshold on a small streak of positives;
(ii)~the lower (recall) threshold is tuned via a two-step heuristic that first identifies an uncorrected empirical threshold $\lo{\tau}_0$, computes normal bounds on the subsets above and below $\lo{\tau}_0$ (scaling both standard errors by the total sample size $\sqrt{s}$ rather than the subset sizes) to form a ``corrected'' recall target, and re-scans for $\lo{\tau}$ without uniform concentration across the candidate grid; and
(iii)~although the pilot sample is drawn from a non-uniform importance distribution biased toward high proxy scores, the precision tuning step in the released code computes unweighted empirical sums of the oracle labels, omitting the importance-sampling weights.

\subsection{SUPG-IT}
SUPG-IT~\citep{liskowski2026streaming} adapts LOTUS's two-threshold tuning algorithm to streaming environments, iteratively refining the cascade thresholds as data batches arrive rather than estimating them in a single pass. 
Although SUPG-IT aims to provide probabilistic guarantees, its statistical foundations inherit and compound the limitations of SUPG and LOTUS. 
Specifically, its bounds rely on asymptotic Gaussian approximations with plug-in variance estimates, employ an ad-hoc ``target clipping'' heuristic to bypass extreme values in small samples, and repeatedly apply fixed-sample confidence bounds to an adaptive, sequential thresholding process without time-uniform corrections for continuous monitoring.

\subsection{ScaleDoc}
ScaleDoc~\citep{zhang2026scaledoc} tunes a two-threshold cascade to meet a target F1~score using proportional stratified sampling and histogram smoothing, and claims a finite-sample $1 - \delta$ guarantee based on Bernstein's inequality (see their Proposition~1). 
However, the theoretical analysis contains several statistical flaws: the bound fails to correct for multiple comparisons across the threshold search space, applies standard i.i.d.\ concentration inequalities to non-i.i.d.\ stratified samples, analyzes raw empirical indicator means whereas the actual algorithm relies on smoothed (and jittered) score histograms, and treats population variances as known without accounting for the uncertainty of variance estimation. 
Furthermore, neither the threshold-selection algorithm (Algorithm~2 in \citet{zhang2026scaledoc}) nor the released code takes $\delta$ as an input or subtracts the derived Bernstein safety margin $\epsilon$; instead, the authors informally argue that the one-bin discretization step in their search acts as a conservative buffer to cover $\epsilon$, even though a fixed bin width does not scale with sample size, label variance, or $\delta$.

\subsection{Stretto}
Stretto~\citep{sanmartino2026stretto} considers a broader space of filter execution plans---jointly tuning model size, KV-cache compression ratio, and cascade thresholds---by optimizing execution cost over a continuous relaxation of the plan space via gradient descent, subject to precision and recall constraints. 
Noting that frequentist confidence bounds are difficult to construct over a large plan space, the authors use marginal Bayesian credible intervals to lower-bound precision and recall. 
However, the use of marginal Bayesian credible intervals does not circumvent the need to correct for adaptive selection over the search space: evaluating the final execution plan on the same sample used for gradient-based optimization introduces optimism (selection) bias into the resulting guarantees.

\subsection{Clustering-Sampling-Voting (CSV)}
Departing from model cascades, CSV~\citep{hou2026linear} groups tuples in an embedding space, queries the oracle on a sampled subset of each cluster, and assigns labels to the remaining items via majority voting, dynamically triggering recursive re-clustering when sampled labels within a cluster exhibit high variance. 
Beyond only aiming to bound per-cluster accuracy rather than dataset-wide precision or recall, its theoretical analysis contains several statistical errors: the proofs misapply fixed-sample concentration inequalities to random stopping times, ignore the adaptivity of recursive re-clustering (which invalidates the bounds by conditioning sub-populations on previously observed labels), and omit a union bound to control the global failure probability across the partitioned dataset.

\fi

\end{document}